\documentclass[%
 reprint,
 amsmath,amssymb,
 superscriptaddress,
 aps
]{revtex4-2}

\usepackage{adjustbox}
\usepackage{enumitem} % in your preamble
\usepackage{graphicx} % Include figure files
\usepackage{svg} % Allow inclusion of SVG images
\usepackage{dcolumn} % Align table columns on decimal point
\usepackage{bm} % bold math
\usepackage{enumerate}
\usepackage{academicons}
\usepackage{mathtools}
\usepackage{physics}
\usepackage{orcidlink}
\usepackage{soul}

\makeatletter
\let\MYcaption\@makecaption
\makeatother
\usepackage{subcaption}
\makeatletter
\let\@makecaption\MYcaption
\makeatother
\usepackage{hyperref}

\usepackage{xcolor}

\begin{document}

\title{Regression of Suspension Violin Modes\\in KAGRA O3GK Data with Kalman Filters}

\author{Lucas Moisset\orcidlink{0009-0001-7357-5572}}
\affiliation{ISAE-ENSMA, Téléport 2, 1 Av. Clément Ader, 86360 Chasseneuil-du-Poitou, France}
\affiliation{ONERA, 2 Av. Marc Pélegrin, 31400 Toulouse, France}

\author{Marco Meyer-Conde\orcidlink{0000-0003-2230-6310}}
\email{contact: marco@tcu.ac.jp}
\email{marco.meyer@cern.ch}
\affiliation{Department of Design and Data Science and Research Center for Space Science, Advanced Research Laboratories, \\ Tokyo City University, 3-3-1 Ushikubo-Nishi, Tsuzuki-ku, Yokohama, Kanagawa 224-8551, Japan}%
\affiliation{University of Illinois at Urbana-Champaign, Department of Physics, Urbana, Illinois 61801-3080, USA}

\author{Christopher Alléné\orcidlink{0009-0001-3859-5420}}
\affiliation{Department of Design and Data Science and Research Center for Space Science, Advanced Research Laboratories, \\ Tokyo City University, 3-3-1 Ushikubo-Nishi, Tsuzuki-ku, Yokohama, Kanagawa 224-8551, Japan}

\author{Yusuke Sakai\orcidlink{0000-0001-8810-4813}}
\affiliation{Department of Design and Data Science and Research Center for Space Science, Advanced Research Laboratories, \\ Tokyo City University, 3-3-1 Ushikubo-Nishi, Tsuzuki-ku, Yokohama, Kanagawa 224-8551, Japan}

\author{Dan Chen\orcidlink{0000-0003-1433-0716}}
\affiliation{National Astronomical Observatory of Japan
2-21-1 Osawa, Mitaka, Tokyo 181-8588, Japan}

\author{Nobuyuki Kanda\orcidlink{0000-0001-6291-0227}}
\affiliation{Department of Physics, Osaka Metropolitan University, Japan}
\affiliation{Nambu Yoichiro Institute of Theoretical and Experimental Physics (NITEP), Osaka Metropolitan University, Japan}

\author{Hirotaka Takahashi\orcidlink{0000-0003-0596-4397}}
\affiliation{Department of Design and Data Science and Research Center for Space Science, Advanced Research Laboratories, \\ Tokyo City University, 3-3-1 Ushikubo-Nishi, Tsuzuki-ku, Yokohama, Kanagawa 224-8551, Japan}%
\affiliation{Earthquake Research Institute, The University of Tokyo, 1-1-1 Yayoi, Bunkyo-ku, Tokyo 113-0032, Japan}

\date{\today}

%%%%%%%%%%%%%%%%%%%%%%%%%%%%%%%%%%%%%%%%
% \begin{figure*}[htb]
%     \centering
%     \begin{tabular}{cc}
%       \begin{minipage}[t]{0.49\textwidth}
%         \centering
%         \includegraphics[width=0.98\textwidth]{xxx.png}
%         \subcaption{`he3.5` model}
%       \end{minipage} &
%       \begin{minipage}[t]{0.49\textwidth}
%         \centering
%         \includegraphics[width=0.98\textwidth]{xxx.png}
%         \subcaption{`y20` model}
%       \end{minipage}
%     \end{tabular}
% \caption{(left) xxx; (right)}
% \label{fig:xxx}
% \end{figure*}
%%%%%%%%%%%%%%%%%%%%%%%%%%%%%%%%%%%%%%%%

%%%%%%%%%%%%%%%%%%%%%%%%%%%%%%%%%%%%%%%%
% \begin{figure}[tb]
%     \centering
%     \includegraphics[width=0.98\columnwidth]{zzz.pdf}
%     \caption{Lorem ipsum color sit amet.}
%     \label{fig:yyy}
% \end{figure}
%%%%%%%%%%%%%%%%%%%%%%%%%%%%%%%%%%%%%%%%

\begin{abstract}
Suspension {thermal} modes in interferometric gravitational-wave detectors produce narrow, high-$Q$ spectral lines that can contaminate {gravitational} searches and bias parameter estimation. In KAGRA, {cryogenic mirrors are held by thick suspension fibers, designed to sustain such a low-temperature environment, which may further affect inharmonicity modes, fiber dimensions, and mechanical behavior compared to typical interferometers. {As these modes remain a prominent source of narrowband contamination, }w}e implement a Kalman filter to model and {track} violin lines, building on the methodology introduced in~\cite{PhysRevD.63.062004}, and apply subtraction to {KAGRA O3GK data}. Using gravitational-wave template injections, we validate that the subtraction preserves matched-filter SNR while effectively suppressing line power. Comparisons of power spectral densities and residual analyses confirm that the method removes deterministic line contributions without introducing waveform distortions. This approach provides a cleaner strain channel for searches and parameter estimation and will become increasingly important for future low-temperature detectors with higher-$Q$ suspensions, such as the Einstein Telescope.
\end{abstract}

\keywords{}

\maketitle

%%%%%%%%%%%%%%%%%%%%%%%%%%%%%%%%%%%%%%%%
\section{Introduction}\label{sec:intro}
%%%%%%%%%%%%%%%%%%%%%%%%%%%%%%%%%%%%%%%%
In KAGRA~\cite{10.1093/ptep/ptac093,10.1093/ptep/ptaa125}, the mirrors {operate} at cryogenic temperatures to reduce thermal noise from the test masses and suspension system. The mechanical $Q$-factor of the sapphire suspension fibers is theoretically expected to be of order $10^7$, whereas experimental measurements in KAGRA report values closer to $10^4$~\cite{10.1093/ptep/ptac093} and also confirmed in Table~\ref{tab:violin-sidebyside}. Such high $Q$ values, whether intrinsic or effective, lead to violin modes that are extremely narrow in frequency, producing tall resonant lines in the strain spectrum. These sharp spectral features contaminate localized frequency regions and must be carefully identified and mitigated to avoid reducing the sensitivity of gravitational-wave searches. Traditional mitigation strategies, such as notch filtering, remove frequency bins around the violin modes. Although effective at suppressing line power, these filters also eliminate any overlapping gravitational-wave signal and introduce phase discontinuities and ringing artifacts. This can distort matched-filter signal-to-noise ratio (SNR) calculations and bias waveform consistency tests such as the $\chi^2$ test~\cite{Allen2005,ref:Allen2012,ref:Creighton_2011,ref:Maggiore_2008}. \\

An adaptive filtering technique based on state-space modeling was pioneered in GW field by \textit{Finn et al.}~\cite{PhysRevD.63.062004,PhysRevD.67.109902}. By representing each violin mode as a damped harmonic oscillator in a discrete-time state-space framework, one can track its dynamics—amplitude and phase—in real time, predict its evolution, and subtract it from the strain channel while consistently accounting for both the physical process and measurement noise. In this work, we implement a Kalman filter to reduce suspension violin modes from KAGRA O3GK public strain data~\footnote{{https://gwosc.org/O3/O3GK/ (Accessed August 2025)}}. The filter is applied to long stretches of data, including segments with injected gravitational-wave signals, to validate its performance. We assess its impact on matched-filter SNR, power spectral densities (PSDs), and residual waveforms, and compare results across different scenarios, including validation tests using the LIGO 40\,m interferometer. This study is the first demonstration of extracting suspension thermal noise at KAGRA in the cryogenic regime and shows that the technique effectively suppresses narrow spectral lines while preserving gravitational-wave signals. 

The paper is structured as follows: Section~\ref{sec:experimental-setup} describes the experimental setup and suspension characteristics; Section~\ref{sec:kalman} details the Kalman filter design and state-space formulation; Section~\ref{sec:result} presents results including LIGO 40\,m validation, KAGRA thermal-noise subtraction, and injection tests; Section~\ref{sec:concl} concludes with discussions on future applications for cryogenic detectors.

%%%%%%%%%%%%%%%%%%%%%%%%%%%%%%%%%%%%%%%%
\section{Experimental Setup}\label{sec:experimental-setup}
%%%%%%%%%%%%%%%%%%%%%%%%%%%%%%%%%%%%%%%%

\begin{figure}[!htb]
    \centering
    \includegraphics[width=\columnwidth]{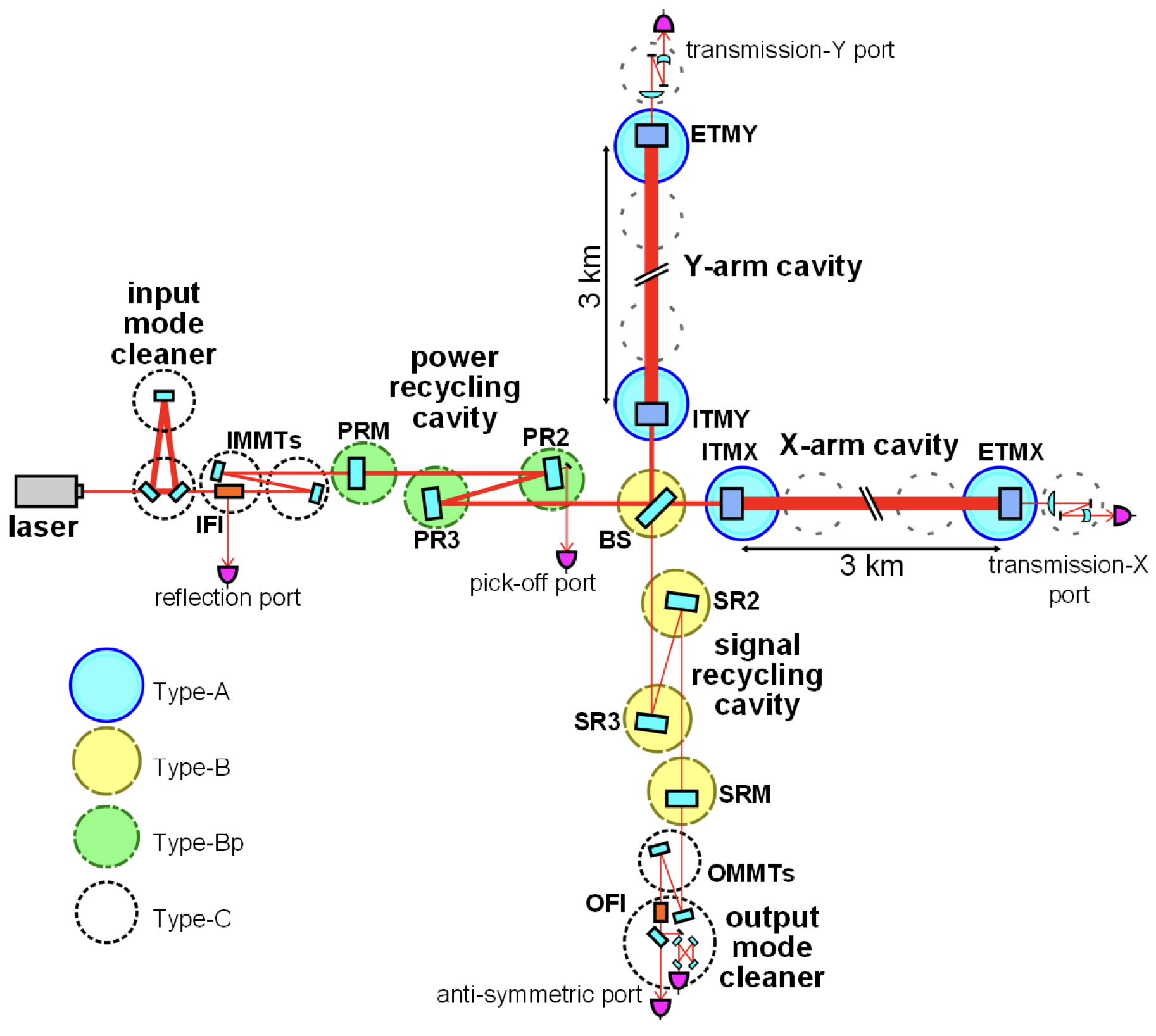}
    \caption{Schematic of the KAGRA interferometer~\cite{10.1093/ptep/ptaa125}. All mirrors with labels are suspended inside the vacuum tanks with four types of vibration isolation systems. The different types of circles in the figure represent the various types of vibration isolation systems. Vacuum tanks in front of the input and end test masses (depicted as dotted grey circles) contain narrow-angle baffles and optical systems for the photon calibrator. ITMX (Y): input test mass X (Y), ETMX (Y): end test mass X (Y), BS: beam splitter, PRM: power recycling mirror, SRM: signal recycling mirror, IMMT (OMMT): input (output) mode-matching telescope, IFI (OFI): input (output) Faraday isolator.}
    \label{fig:kagra-ifo}
\end{figure}

This work focuses on suppressing suspension violin modes in interferometric gravitational-wave detectors, with particular emphasis on the KAGRA cryogenic interferometer. While the full interferometer includes multiple suspended optics, our analysis concentrates on the type-A test mass suspensions, circled in light blue in Fig.~\ref{fig:kagra-ifo}, which are the dominant sources of narrow, high-$Q$ violin modes in the DARM channel. Since the DARM channel is both the primary observable for gravitational waves and the only channel available to monitor these suspension noises, subtraction is challenging without potentially affecting astrophysical signals. This makes the Kalman filter an attractive approach, as it enables real-time noise estimation and removal without introducing external or potentially disruptive elements that could compromise the stability of such low-temperature systems.

\begin{figure}[!htb]
    \centering
    \includegraphics[width=\columnwidth]{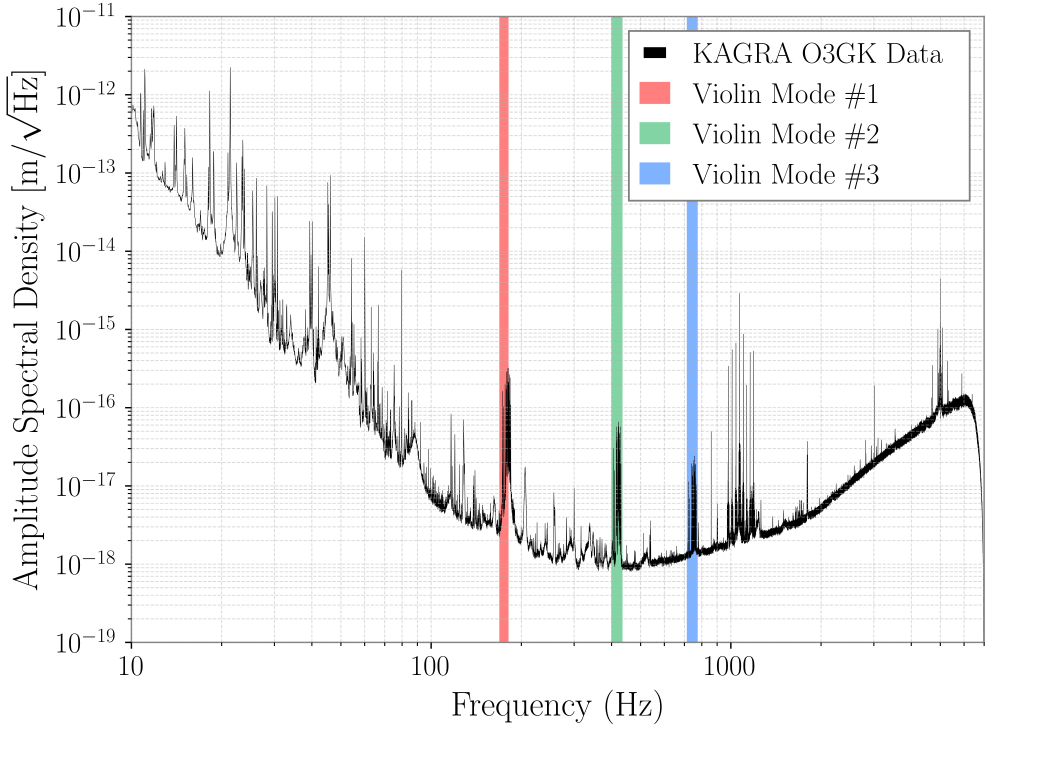}
    \caption{{O3GK strain sensitivity at GPS 1270312960, highlighting the first three violin modes, which are the most prominent suspension thermal noise and the focus of this work.}}
    \label{fig:KAGRA-PSD-O3GK}
\end{figure}

The PSD of the KAGRA O3GK data highlights the region of the violin-mode noise in Fig.~\ref{fig:KAGRA-PSD-O3GK}. This suspension thermal noise appears in a particularly critical frequency band of the O3GK data.

%%%%%%%%%%%%%%%%%%%%%%%%%%%%%%%%%%%%%%%
%%%%%%%%%%%%%%%%%%%%%%%%%%%%%%%%%%%%%%%

%%%%%%%%%%%%%%%%%%%%%%%%%%%%%%%%%%%%%%%%
\subsection{Suspension Design at KAGRA}\label{subsec:kagra-setup}
%%%%%%%%%%%%%%%%%%%%%%%%%%%%%%%%%%%%%%%%

\begin{figure}[!htb]
    \centering
    \includegraphics[width=0.95\columnwidth]{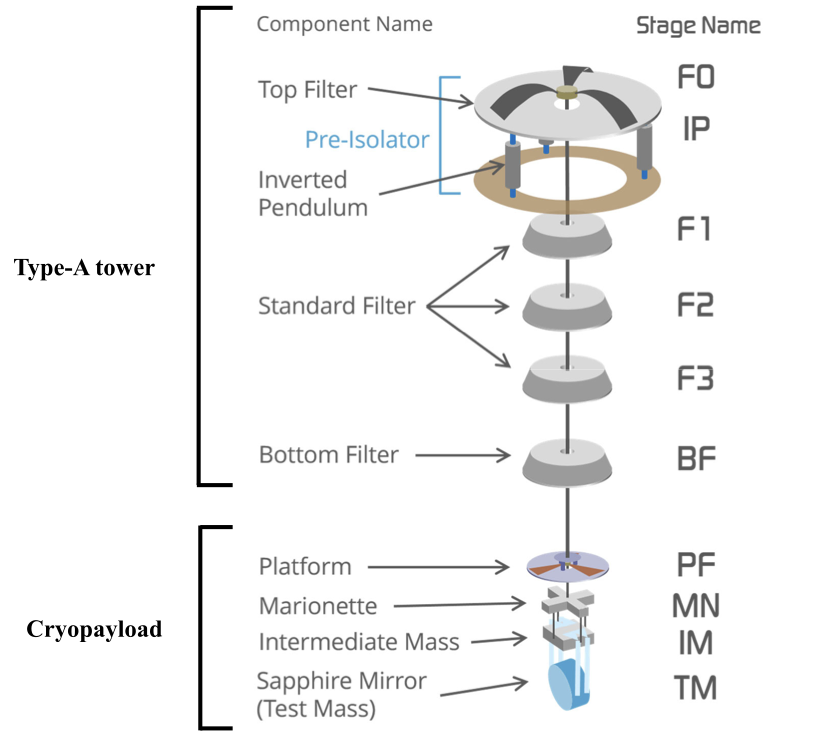}
    \caption{Schematic of Type-A suspension~\cite{10.1093/ptep/ptaa125}. The suspension is a multiple-stage pendulum with a total height of 13m. The Type-A tower is at room temperature, while the cryopayload is cooled to cryogenic temperatures.}
    \label{fig:type-a}
\end{figure}

KAGRA employs a multi-stage pendulum suspension system to isolate its test masses from seismic and thermal noise, shown in Fig.~\ref{fig:type-a}. The final stage consists of a sapphire test mass suspended by sapphire fibers, operated at cryogenic temperatures to reduce thermal fluctuations. The high quality factor ($Q \gtrsim 10^7$) of these fibers results in narrow, long-lived resonances — known as violin modes — {in the 170-186\,Hz frequency range}, along with higher harmonics.

%%%%%%%%%%%%%%%%%%%%%%%%%%%%%%%%%%%%%%%
%%%%%%%%%%%%%%%%%%%%%%%%%%%%%%%%%%%%%%%

\begin{figure}[!htb]
    \centering
    \includegraphics[width=\columnwidth]{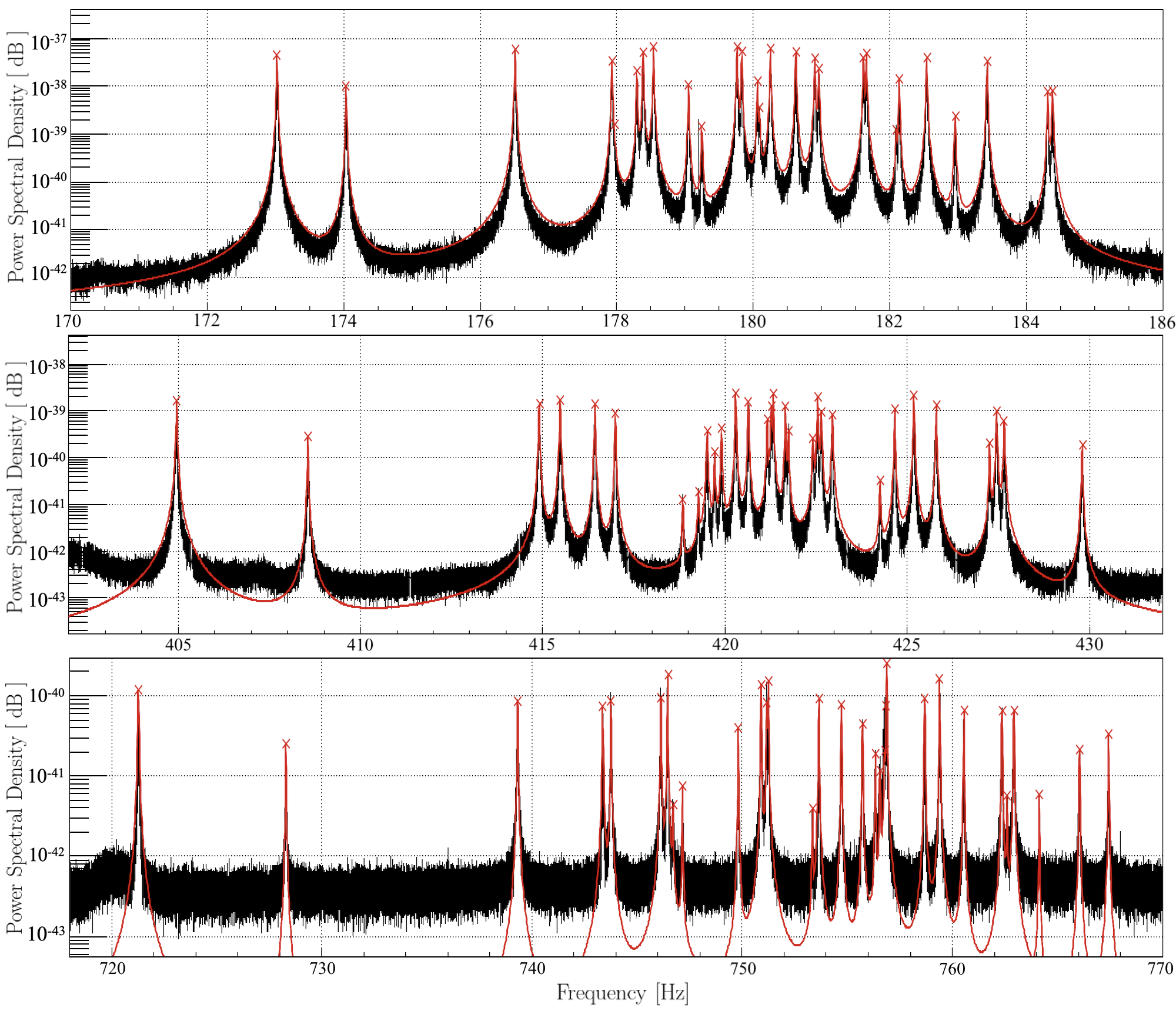}
    \caption{The estimated PSD (GPS 1271267328..1271361536) in the first three violin-mode bands is shown in black, with the fitted Lorentzian profiles overlaid in red. (a) Top: first mode between 170-186Hz ($\Delta f_0 \approx 16$\,Hz); (b) Middle: second mode ranging from 402-432Hz ($\Delta f_1 \approx 30$\,Hz); (c) Bottom: third modes ranging from 718-770Hz ($\Delta f_2 \approx 52$\,Hz)}
    \label{fig:violin-modes}
\end{figure}

Fig.~\ref{fig:violin-modes} shows the power spectral density (PSD) in the DARM channel. {The PSD is modeled as a sum of separately fitted Lorentzian distributions, one for each of the $N$ identified peaks}

\begin{equation}
S_x(f) = \sum_{i=1}^{N} \frac{G_i \,f_{0,i}^2\Gamma_i^2}{\left(f^2 - f_{0,i}^2\right)^2 + f^2\Gamma_i^2}
\label{eq:lorentz_sum_hwhm}
\end{equation}
where
\begin{itemize}
\item $G_i$ is the amplitude of the $i$-th peak,
\item $f_{0,i}$ is the central frequency of the $i$-th peak,
\item $\Gamma_i$ is the full-width at half maximum (FWHM) of the $i$-th peak
\end{itemize}

No background contribution is included in this fit. {Also,} for high-$Q$ resonances, the quality factor of each mode is given by
\begin{equation}
Q_i = \frac{f_{0,i}}{\mathrm{\Gamma}_i}
\label{eq:qi_def_hwhm}
\end{equation}
\vspace{-0.1cm}
%%%%%%%%%%%%%%%%%%%%%%%%%%%%%%%%%%%%%%%
%%%%%%%%%%%%%%%%%%%%%%%%%%%%%%%%%%%%%%%

\begin{table}[htb]
\centering
\caption{A summary of the estimated peak parameters from our Lorentzian fits is provided here, while the complete list of individual resonances and their fitted values is reported in Appendix~\ref{app:frequencies}. {Theoretical Q-values are taken from \cite{10.1093/ptep/ptac093} at room temperature}}
\renewcommand{\arraystretch}{1.10}
\begin{tabular}{lccc}
\hline\hline
 & 1st mode & 2nd mode & 3rd mode \\
\hline
Number of identified peaks & 27 & 30 & 31 \\
Peak frequency (theory)    & 175 Hz & 410 Hz & 732 Hz \\
Peak frequency (mean)      & 180.10 Hz & 420.84 Hz & 752.47 Hz \\
Peak frequency (lowest)    & 173.02 Hz & 404.96 Hz & 721.27 Hz \\
Peak frequency (highest)   & 184.39 Hz & 429.80 Hz & 767.44 Hz \\
$Q$-value (theory)         & $2.7 \times 10^4$ & $4.7 \times 10^4$ & $7.0 \times 10^4$ \\
$Q$-value (mean)           & $1.30 \times 10^4$ & $1.80 \times 10^4$ & $2.12 \times 10^4$ \\
$Q$-value (std.dev.)      & $2.0 \times 10^3$ & $3.7 \times 10^3$ & $6.9 \times 10^3$ \\
\hline\hline
\vspace{-0.4cm}
\end{tabular}
\label{tab:violin-mode-parameters}
\end{table}
In cryogenic conditions, the coupling of thermal noise can {significantly} differ from that in room-temperature suspensions. This makes its suppression particularly relevant for KAGRA, especially given the discrepancy between theoretical expectations and experimental measurements reported in \cite{10.1093/ptep/ptac093}. Furthermore, the violin-mode resonances do not align with exact integer multiples of the fundamental frequency. In particular, the second and third harmonics deviate from {the usual harmonic scaling expected for a thin fiber}, as exhibited by {the} LIGO40m setup in Fig.~\ref{fig:ligo40m-psd-modes}. This deviation illustrates that the sapphire suspension wires in KAGRA are subject to material rigidity effects~\cite{10.1121/1.1906888}, which give rise to inharmonicity, in close analogy to what has been observed in the LIGO and Virgo experiments. In the evaluation of the first three mode peaks, we observed that only 27, 30, and 31 peaks were identified, as shown in Fig.~\ref{fig:violin-modes} and Table.~\ref{tab:violin-mode-parameters}. Faint or missing peaks hinder direct estimation of the $Q_i$ factor, while two closely spaced resonances may blend into a single feature, leading to an artificially inflated $Q$ value. This effect is examined in detail in the following section.

%%%%%%%%%%%%%%%%%%%%%%%%%%%%%%%%%%%%%%%%
\subsection{Second order resonant effects}\label{subsec:yaw-pitch-roll}
%%%%%%%%%%%%%%%%%%%%%%%%%%%%%%%%%%%%%%%%

\begin{figure}[!htb]
    \centering
    \includegraphics[width=0.65\columnwidth]{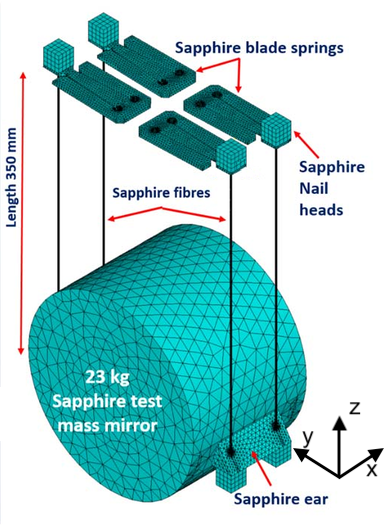}
    \caption{The CAD model of the mirror and its suspension wires~\cite{JoPCS.10.1088}. The $x$-axis corresponds to the interferometer’s sensitive direction, while modifications along the $y$ and $z$ axes are effectively shadowed and therefore undetectable.
}
    \label{fig:mirror-cad}
\end{figure}

The CAD model of the mirror and its suspension wires is shown in Fig.~\ref{fig:mirror-cad}.
In addition to the fundamental violin modes of the suspension fibers along {the beam axis, {denoted $x$}, a} coupling from higher-order resonances and other degrees of freedom - such as yaw, pitch, and roll rotations around the mirror’s center of mass - may introduce secondary spectral features. Further details can be found in Appendix \ref{app:so-formula}. In a first-order angular development, only the yaw contribution (Oz) is expected to be significant, since the lever arm is set by approximately half the mirror diameter along the y-axis for each suspension wire. The pitch degree of freedom is neglected here under the assumption that the attachment points of the fibers are coplanar in the $xOy$ plane; if the wires were not perfectly coplanar, vertical offsets would introduce additional pitch coupling. The roll degree of freedom is neglected at first order because displacements along the $x$-axis are largely insensitive to rotations around this axis.

%%%%%%%%%%%%%%%%%%%%%%%%%%%%%%%%%%%%%%%
%%%%%%%%%%%%%%%%%%%%%%%%%%%%%%%%%%%%%%%
\vspace{0.5cm}

\begin{figure}[!htb]
    \centering
    \includegraphics[width=\columnwidth]{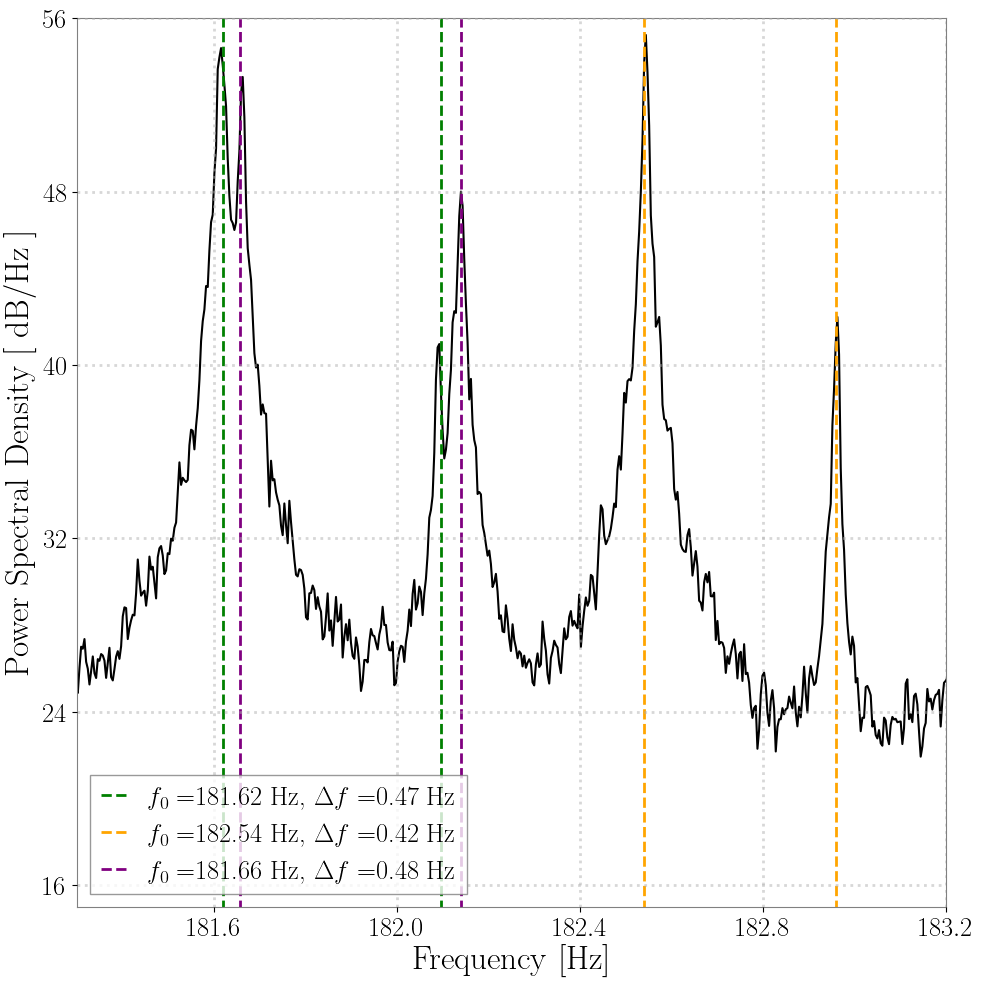}
    \caption{High-resolution PSD computed using KAGRA data. {Each of the longitudinal-direction modes is expected to have a companion mode at a higher frequency due to yaw coupling—see the pairs highlighted in green, purple, and orange. The candidate} shifts of about $0.45\,\mathrm{Hz}$ {are} consistent with the theoretical lower bound estimated in Appendix~\ref{app:so-formula}.}
    \label{fig:psd-highresolution}
\end{figure}

In practice, these second-order couplings can produce weak sidebands or nearby peaks in the power spectral density (PSD) at frequency offsets $\Delta f_0$, complicating the filtering process in addition to the longitudinal-like resonance strongly coupled with the arm length's readout. Auxiliary peaks, sharing a common $Q_i$ value, typically have much lower amplitudes as {they couple} through geometric asymmetries. To illustrate such {an} effect, we computed a high-resolution PSD in Fig.~\ref{fig:psd-highresolution} similar to {that in} Finn \textit{et al.}, highlighting the secondary lower-amplitude yaw-like peak candidates. Some lower amplitude peaks may be buried in the measurement noise or degenerate with other resonances, making them less visible in the PSD distribution. Since the measurement noise is lower for the second and third modes, it is easier to perform proper peak extraction in this frequency region; explaining why only 27 peaks are found in the first mode, as reported in Table~\ref{tab:violin-mode-parameters}.

%%%%%%%%%%%%%%%%%%%%%%%%%%%%%%%%%%%%%%%

%%%%%%%%%%%%%%%%%%%%%%%%%%%%%%%%%%%%%%%

%%%%%%%%%%%%%%%%%%%%%%%%%%%%%%%%%%%%%%%%
\section{Kalman Filter Design}\label{sec:kalman}
%%%%%%%%%%%%%%%%%%%%%%%%%%%%%%%%%%%%%%%%

We employ a discrete-time Kalman filter to model, infer, and subtract violin-mode contributions from the interferometer strain channel. The filter is designed to operate on long stretches of data, making it suitable for online or low-latency subtraction. The Kalman filter provides a time-discrete, recursive estimate of the suspension wire’s resonant motion. Given prior knowledge (state-space model and noise covariances), an initial state, and measurements sampled at frequency $f_s$, it alternates prediction and update steps to infer the latent state at each sample, reconstructing the true motion in the presence of process and measurement noise. 

%%%%%%%%%%%%%%%%%%%%%%%%%%%%%%%%%%%%%%%%
\subsection{Suspension thermal dynamics}\label{subsec:thermal-dynamics}
%%%%%%%%%%%%%%%%%%%%%%%%%%%%%%%%%%%%%%%%
The dynamics of a single violin mode are modeled as a damped harmonic oscillator with resonance frequency $\omega_0$ and quality factor $Q${, described by the equation of motion.}
\begin{equation}
m\ddot{x}(t) + b\,\dot{x}(t) + k\,x(t) = 0
\end{equation}
where $k$ and $b$ are the stiffness and viscous damping coefficients of the suspension fiber, respectively.

For each suspension wire, the natural frequency and quality factor are defined as
\begin{equation}
\omega_0 = \sqrt{\frac{k}{m}}, \qquad Q = \frac{m\omega_0}{b}
\end{equation}

In our setup, the stochastic dynamics are described by the Langevin model:
\begin{equation}
\ddot{x}(t) + \frac{\omega_0}{Q}\,\dot{x}(t) + \omega_0^2 x(t) = \dfrac{1}{m}\xi(t)
\end{equation}
where $x(t)$ is the displacement along the oscillation axis and $\xi(t)$ is the zero-mean stochastic Langevin force {with statistical properties}
\begin{equation}
\langle \xi(t) \rangle = 0, \qquad 
\langle \xi(t)\,\xi(t') \rangle = 2 \lambda k_B T \, \delta(t-t')
\end{equation}
{where $k_B$ is the Boltzmann constant, $T$ is the temperature of the fiber, and $\lambda$ is the damping coefficient related $Q$-value of the oscillator.}

In accordance with the fluctuation--dissipation theorem
\cite{PhysRev.83.34}, the single-sided spectral density of the Langevin
force, $S_F(\omega)$, is determined by the dissipative part of the
mechanical response. Explicitly,
\begin{equation}
S_F(\omega) = 4 k_B T \, \Re\!\left[ Z(\omega) \right]
\end{equation}
where $Z(\omega)$ is the mechanical impedance defined above. This
relation shows that thermal driving of the oscillator is entirely fixed
by its energy dissipation properties, linking microscopic Langevin noise
to the macroscopic damping encoded in $\Re[Z(\omega)]$.

%%%%%%%%%%%%%%%%%%%%%%%%%%%%%%%%%%%%%%%%
\subsection{Slowly varying envelope approximation}\label{subsec:dae}
%%%%%%%%%%%%%%%%%%%%%%%%%%%%%%%%%%%%%%%%

{If} the high-frequency oscillations of the violin mode decay on a timescale much shorter than the frequency scale of interest in the Laplace domain ($|p_+| \ll |s|$, with $p_+$ denoting the first pole and $s$ the Laplace variable){, the} inertial term $\ddot{x}(t)$ can {be averaged away}, {and therefore} the second-order oscillator dynamics reduces to an effective first-order equation governing the slow envelope of the motion. Within this approximation, the displacement can be expressed as a rapidly oscillating carrier modulated by a slowly varying complex amplitude:
\begin{equation}
x(t) \approx \tilde{x}(t) \, e^{i \omega_0 t} + \tilde{x}^*(t) \, e^{-i \omega_0 t}
\end{equation}
where $\tilde{x}(t)$ evolves slowly compared to $e^{\pm i \omega_0 t}$ ( $\ddot{x} \ll \dot{x}$ ). Substituting into the oscillator equation yields
\begin{equation}
\frac{\omega_0}{Q} \, \dot{x}(t) + \omega_0^2 x(t) = \xi(t)
\end{equation}
which captures the slow dynamics of the envelope while retaining the oscillatory carrier. The solution to this equation is:
\begin{equation}
p_{\pm} = -\frac{\omega_0}{2Q} \pm i \omega_0
\end{equation}

This first-order approximation allows the violin mode to be modeled directly in a linear Kalman filter, avoiding the overhead of an extended filter while still reproducing the dominant thermal-driven ringdown relevant for mode subtraction.

%%%%%%%%%%%%%%%%%%%%%%%%%%%%%%%%%%%%%%%%
\subsection{Data conditioning}\label{subsec:data-conditioning}
%%%%%%%%%%%%%%%%%%%%%%%%%%%%%%%%%%%%%%%%

Before applying the Kalman filter, the measured signal $x(t)$ is shifted in time and only then downsampled to match the narrow bandwidth $\Delta f_c$ of the targeted mode per the method of \textit{Finn et al.}\cite{PhysRevD.63.062004}. This ensures that only the selected spectral band is affected, preventing unintended corrections outside the mode while preserving the broadband strain.

To isolate the mode dynamics, the signal is first transformed into a complex narrowband form around the target frequency $f_c$ (\textit{i.e.} $\omega_c = 2\pi f_c$) :
\begin{equation}
\psi(t) = x(t) \, e^{-i \omega_c t}, 
\qquad \omega_c = 2 \pi f_c
\end{equation}
where the real and imaginary parts of $\psi(t)$ correspond to the in-phase and quadrature components of the mode.

Introducing the vector notation, the dynamics are now expressed in a form suitable for a linear Kalman filter. 
\begin{equation}
\boldsymbol{\psi}(t) =
\begin{pmatrix}
\Re[\psi(t)] \\
\Im[\psi(t)]
\end{pmatrix},
\qquad
\dot{\boldsymbol{\psi}}(t) = \boldsymbol{A} \, \boldsymbol{\psi}(t) + \boldsymbol{\xi}(t)
\end{equation}
\begin{equation}
\boldsymbol{A} \approx
\begin{pmatrix}
- \dfrac{\omega_0}{2 Q} & \, \omega_c - \omega_0 \\
\omega_0 - \omega_c & - \dfrac{\omega_0}{2 Q}
\end{pmatrix}
\end{equation}
This representation provides the continuous-time formulation, which can then be discretized according to the sampling rate $\Delta f_c$ as $\boldsymbol{A'} = \exp(\boldsymbol{A}/\Delta f_c)$. After processing with the Kalman filter, the narrowband estimate of the violin mode, $\boldsymbol{\hat{\psi}}_k$, is upsampled to the original sampling rate and rotated back to the original frequency:
\begin{equation}
x_{\rm v}(t) = 2 \, \Re \Big[ \hat{\psi}(t) \, e^{i \omega_c t} \Big]
\end{equation}

The corrected real-valued strain signal is then obtained by subtracting the estimated mode contribution:
\begin{equation}
x_{\rm_{corr}}(t) = x(t) - x_{\rm v}(t)
\end{equation}

This completes the forward and reverse processing steps, allowing the violin-mode contribution to be removed while preserving the broadband interferometer strain.

%%%%%%%%%%%%%%%%%%%%%%%%%%%%%%%%%%%%%%%%
\subsection{Observation and state-space representation}\label{subsec:state-space}
%%%%%%%%%%%%%%%%%%%%%%%%%%%%%%%%%%%%%%%%
The discrete-time Kalman filter is defined by the standard state-space equations:
\begin{equation}
\left\{
\begin{aligned}
\boldsymbol{\psi}_{k+1} &= \mathbf{A}' \, \boldsymbol{\psi}_k + \mathbf{w}_k \\
\mathbf{z}_k &= \mathbf{C} \, \boldsymbol{\psi}_k + \mathbf{v}_k
\end{aligned}
\right.
\label{eq:kalman_system}
\end{equation}
\begin{equation}
\mathbf{w}_k \sim \mathcal{N}(\mathbf{0}, \mathbf{W}), \quad
\mathbf{v}_k \sim \mathcal{N}(\mathbf{0}, \mathbf{V})
\label{eq:noise_def}
\end{equation}
where,
\begin{itemize}
\item $\boldsymbol{\psi}_k \in \mathbb{R}^{2N}$ contains the in-phase ($\mathfrak{R}$) and quadrature ($\mathfrak{I}$) components for each targeted mode,
\item $\mathbf{z}_k \in {\mathbb{R}^2}$ is the discrete observation vector derived from the complex-demodulated strain,
\item $\mathbf{A'}$ is the discrete-time state transition matrix, derived from the oscillator parameters $\omega_c$, $\omega_0$, and $Q$ for each mode,
\item $\mathbf{C}$ maps the state vector to the measurement vector,
\item $\mathbf{w}_k$ and $\mathbf{v}_k$ are zero-mean process and measurement noise vectors {with their} covariance matrices {denoted} $\mathbf{W}$ and $\mathbf{V}${, respectively}.
\end{itemize}

In practice, the filter is applied directly to measured interferometer data; no simulated data are generated. However, the same equations can also be used to generate sample data that follow the system dynamics as defined by $\mathbf{A}'$, $\mathbf{Q}$, and $\mathbf{R}$, which is useful for testing or validation purposes.\\

Some peaks lie within a few mHz; it is therefore necessary to apply a joint Kalman filter extraction along the line{s of} Finn \textit{et al.} \cite{PhysRevD.63.062004,PhysRevD.67.109902}. Such extraction is performed by defining the new matrix elements as :
\begin{align*}
\mathbf{A} &= \mathrm{diag}(\mathbf{A}_1, \dots, \mathbf{A}_N), &
\mathbf{C} &= 
\begin{bmatrix}
\mathbf{I}_2 & \mathbf{I}_2 & \cdots & \mathbf{I}_2
\end{bmatrix}, \\
\mathbf{W} &= \mathrm{diag}(\mathbf{W}_1, \dots, \mathbf{W}_N), &
\mathbf{V} &\propto \mathbf{I}_2
\end{align*}

The precise {estimation} of the process and measurement noise covariances {is determined by} $\sigma_{w_i}$ and $\sigma_{v}${, which denote} the standard deviations of the $i$-th process noise component and the $j$-th measurement noise component, respectively.
\begin{equation}
\mathbf{W_i} = \sigma_{w,i}^2 \mathbf{I}_n, \quad
\mathbf{V} = \sigma_{v}^2 \mathbf{I}_2
\end{equation}

A definition of the process noise variance $\sigma_{w,i}^2$ and the measurement noise variance $\sigma_{v}^2$ is given in Eq.~\ref{eq:sigmas} {and derived in Appendix} \ref{app:so-formula}.
{
\begin{equation}
\left\{
\begin{aligned}
\sigma_{w,i}^2 &= \dfrac{\omega_{0,i} k_B T}{4 Q^2 \omega_c^2 m \,\Delta f_c} \\[1ex]
\sigma_{v}^2   &= \dfrac{1}{\Delta f_c}\,\dfrac{\bar{P}_m}{2}
\end{aligned}
\right.
\label{eq:sigmas}
\end{equation}}

These quantities are estimated from the physical suspension parameters $(Q, f_0, k, b)$, which are strongly temperature dependent. In particular, cryogenic operation at KAGRA presents a unique opportunity to investigate how the properties of the suspension material evolve with temperature and, consequently, how they impact the noise budget. Measuring these dependencies would be highly valuable for improving violin-mode modeling and subtraction in future observing runs.\\

Note that Finn \textit{et al.}~\cite{PhysRevD.63.062004,PhysRevD.67.109902} defined the process noise within a narrow frequency window of only a few mHz around each resonant peak. In our analysis, we take the peak maximum itself as the reference point for setting the process noise. This choice reflects the fact that several resonances lie very close together, making them difficult to isolate within such narrow bands. Using the peak maximum ensures a well-defined filter response under the narrowband approximation $\omega \approx \omega_0$. 
This provides a consistent prescription for evaluating the process noise through the inversion of the transfer function $\mathbf{H^{-1}}$: 
\begin{equation}
S_F(\omega \approx \omega_0) \;=\; \bigl| H^{-1}(i\omega) \bigr|^2 \, S_x(\omega)
\end{equation}
where $S_F$ denotes the force PSD associated with the Langevin model and $S_x$ the measured displacement PSD around $\omega \approx \omega_0$.

%%%%%%%%%%%%%%%%%%%%%%%%%%%%%%%%%%%%%%%%
\section{Results and Discussions}\label{sec:result}
%%%%%%%%%%%%%%%%%%%%%%%%%%%%%%%%%%%%%%%%

In this section, {we will discuss} the preliminary results of our {work by} simulating a physical true state vector and {observations of it}.

\subsection{Sampling synthetic data}\label{subsec:Sampling}

\begin{figure}[!htb]
    \centering
    \includegraphics[width=\columnwidth]{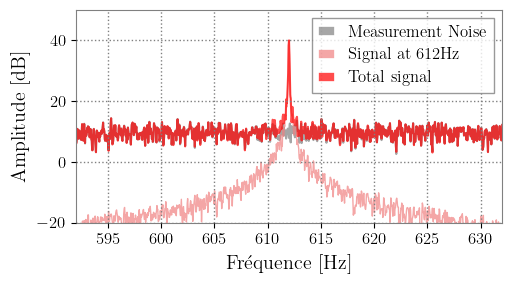}
    \caption{Power spectral density of the PSD noise baseline and the resonant signal {after passing through the} $H(iw)$ complex filter.}
    \label{fig:psd-synthetic}
\end{figure}

The synthetic data {is sampled} by implementing Eqs.~(\ref{eq:kalman_system}) and (\ref{eq:noise_def}) as shown in Fig.~\ref{fig:psd-synthetic}{,} by introducing the $\mathbf{w}_k$ and $\mathbf{v}_k$ at the generation stage. 

In practice, obtaining a reliable estimate of the process noise is crucial: it quantifies stochastic perturbations and unmodeled dynamics that drive the system away from its ideal theoretical trajectory along the time. Keeping this contribution as low as possible reduces systematic biases in the reconstructed wire position, thereby constraining interferometer uncertainties associated with suspension dynamics. By contrast, the measurement noise originates from interferometer uncertainties and readout imperfections, and its level can be characterized from the power spectral density (PSD) floor of the detector. A higher measurement noise implies that the Kalman filter must rely more heavily on the model prediction, effectively increasing the filter’s flexibility in adjusting its resolution of the true state. Conversely, when the measurement noise is low, the filter can afford to weight observations more strongly, leading to a tighter tracking of the measured signal. In interferometric applications, this balance is critical: overestimating measurement noise risks smoothing away real physical fluctuations, while underestimating it can amplify spurious detector noise and degrade the reconstruction.\\

\begin{figure}[!htb]
    \centering
    \includegraphics[width=\columnwidth]{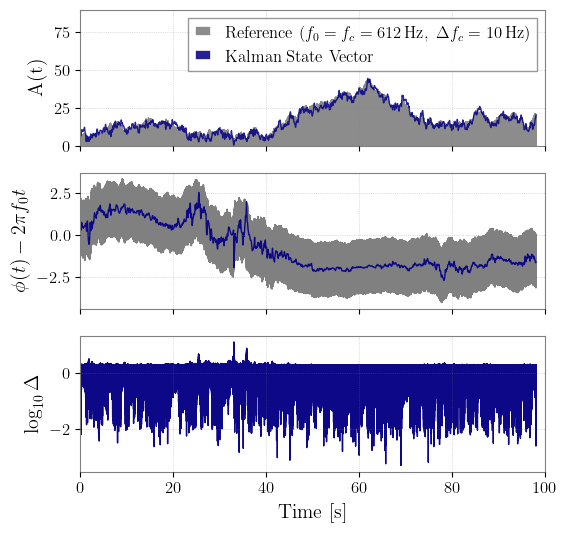}
    \caption{(a) Top: Amplitude {(in dB)} of the reconstructed state vector compared with the noisy true state vector, including process noise. (b) Middle: Unwrapped phase {(in rad)} of the reconstructed state vector, with the linear trend $2\pi f_{0}t$ subtracted to align in time. (c) Bottom: Residual error distribution in logarithmic scale (relative symmetric error).}
    \label{fig:performance-synthetic}
\end{figure}

The inference of the state vector through the Kalman filter allows for a time-dependent reconstruction of the dynamics of a specific suspension as illustrated in Fig.~\ref{fig:performance-synthetic}. At each step, the filter updates its estimate by combining the predicted state from the physical model with the actual interferometer measurements, effectively isolating the resonant motion of the wire and its violin modes from the surrounding detector noise.\\

In Fig.~\ref{fig:performance-synthetic}c, the relative symmetric error {measures} the disparity between the true and reconstructed states and is defined as
\begin{equation}
\Delta(t) = 2\dfrac{|x(t)-\hat{x}(t)|}{|x(t)+\hat{x}(t)|}
\end{equation}
Distinct peaks in $\log_{10}\Delta(t)$ indicate localized mismatches in the reconstruction, as seen after 30–40s. We expect that a Kalman smoother would mitigate these large uncertainties by combining both past (forward pass) and future (backward pass) information. This bidirectional approach reduces transient mismatches and sharp peaks in the relative symmetric error, thereby improving the reconstruction stability. Additionally, the large gray band in Fig.~\ref{fig:performance-synthetic}b reflects the track uncertainty induced by process noise.

\subsection{Reproduction of LIGO 40m noise suppression}\label{subsec:results-ligo40m}

As a validation step, we reproduce the violin-mode subtraction performance demonstrated in the LIGO 40\,m interferometer~\cite{PhysRevD.63.062004}. The 40\,m testbed provides a strong reference for validating our implementation, allowing direct comparison between our implementation and previously established results. 

%%%%%%%%%%%%%%%%%%%%%%%%%%%%%%%%%%%%%%%
\begin{figure}[!htb]
    \centering
    \includegraphics[width=\columnwidth]{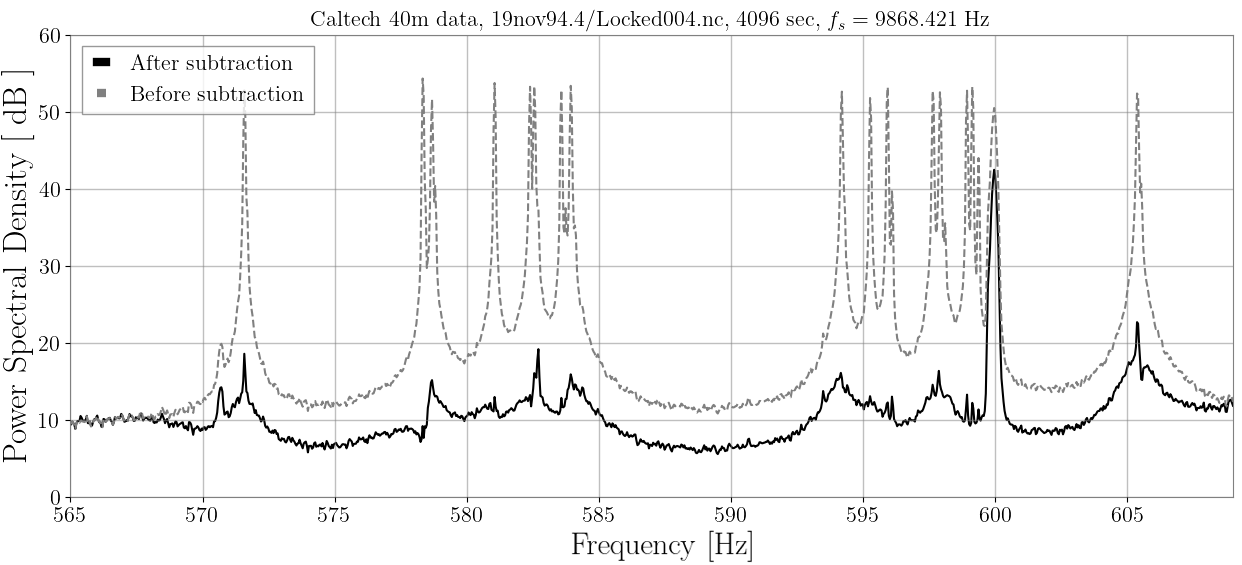}
    \caption{Power spectral density of LIGO 40\,m data. After subtraction (black curve), most of the resonances are suppressed by about $30\,\mathrm{dB}$, which corresponds to a reduction in power by a factor of $10^3$.}
    \label{fig:ligo40m-psd}
\end{figure}
%%%%%%%%%%%%%%%%%%%%%%%%%%%%%%%%%%%%%%%

This first reproduction confirms that our Kalman filter framework reproduces the suppression performance and residual characteristics reported in earlier work from 2001~\cite{PhysRevD.63.062004}. A key feature is the strong reduction of stationary PSD noise in the vicinity of the resonant peaks after Kalman filtering. In our framework, higher-order resonances that lie very close to one another are already modeled and subtracted jointly, ensuring robust suppression. Looking ahead, {we aim} to refine the estimation of the process noise covariance $W$ and the measurement noise covariance $V$ by having a deeper understanding of the underlying hardware system, enabling more optimal suppression. As shown in Fig.~\ref{fig:ligo40m-psd}, the filter also achieves strong suppression (on the order of {10\,dB}) of the 10th harmonic of the {60\,Hz} U.S. power grid. This occurs because both the grid harmonics and the violin resonances are governed by similar differential equations, leading the filter to treat them in a coupled manner. For optimal performance, such additional spectral lines should also be modeled and corrected jointly. However, since the focus of this paper is on the impact of cryogenic operation, their treatment is left to future work.

%%%%%%%%%%%%%%%%%%%%%%%%%%%%%%%%%%%%%%%%
\subsection{Thermal Noise Subtraction using O3GK KAGRA Data}\label{subsec:results-kagra}
%%%%%%%%%%%%%%%%%%%%%%%%%%%%%%%%%%%%%%%%

\begin{figure}[!htb]
    \centering
    \includegraphics[width=0.95\columnwidth]{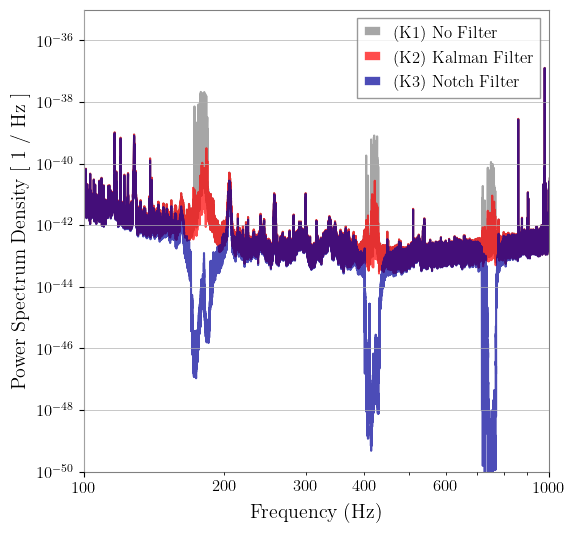}
    \caption{Power spectral density distributions for (K1), (K2), and (K3) cases.}
    \label{fig:kagra-psd-k123}
\end{figure}

\begin{figure}[!htb]
    \centering
    \includegraphics[width=\columnwidth]{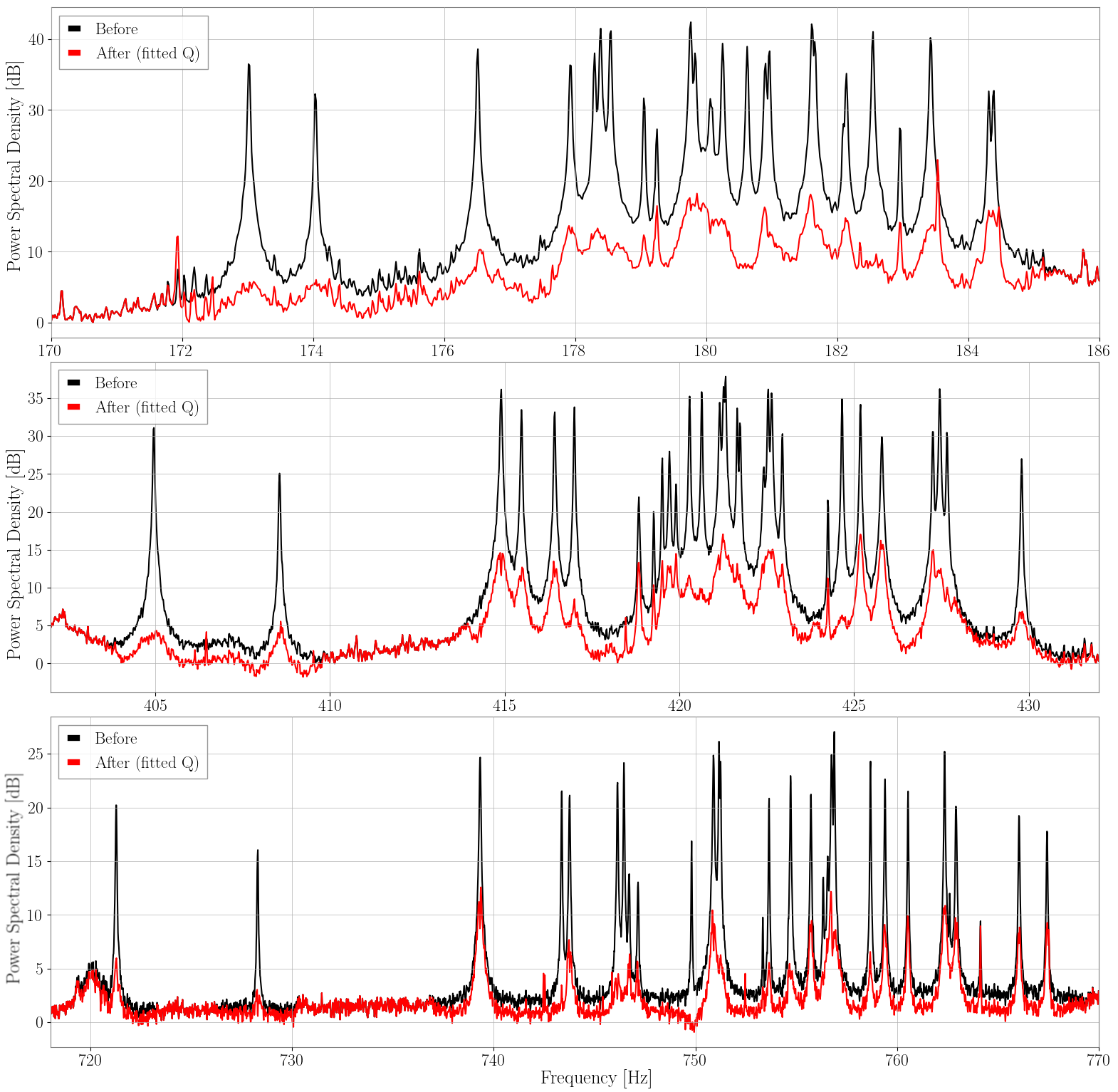}
    \caption{Power spectral density (PSD) of O3GK strain data before and after Kalman filtering. The process and measurement noise parameters are still being tuned for improved subtraction. The vertical scale is in dB (\textit{i.e.} $10\log_{10}(P/P_{\mathrm{ref}})$); the baseline offset is arbitrary and chosen for visual clarity.}
    \label{fig:kagra-psd-kalman}
\end{figure}

Applying the Kalman filters defined in Table~\ref{tab:kalman_bands_fc} to the KAGRA O3GK public data set yields a marked suppression of the targeted violin modes, as illustrated in Fig.~\ref{fig:kagra-psd-kalman}. In practice, the resonance frequency $f_{0}$, the quality factor $Q$, and the mode amplitude $A$ vary slowly with time due to environmental and instrumental fluctuations. The Kalman filter framework is particularly well-suited for this situation, since the process noise covariance $W$ and measurement noise covariance $V$ explicitly account for such variability. By continuously adapting to these fluctuations, the filter is able to track the resonant dynamics in real time and achieve robust subtraction of violin-mode contributions, while preserving the broadband astrophysical signal. The effectiveness of the Kalman filtering is evaluated by considering {the cases shown in Fig.~\ref{fig:kagra-psd-k123}:}

\begin{description}[
    style=unboxed,
    leftmargin=!,
    labelwidth=1cm,
    topsep=5pt,    % space above and below the list
    partopsep=5pt, % extra space when starting a new paragraph
    itemsep=0.1em % space between items
]
  \item[(K1)] SNR computed without Kalman filtering in the first three modes of the frequency band,
  \item[(K2)] SNR after Kalman filtering the three modes,
  \item[(K3)] SNR after violin-mode forest subtraction using an IIR band-stop around the resonant peaks.
\end{description}

Our results show that the filter reduces narrow spectral lines by more than 20\,dB in the targeted frequency bins, without introducing strong distortion to the surrounding broadband noise. {A 3\,dB reduction corresponds to halving the power ($P \to P/2$), \textit{i.e.}, a $1/\sqrt{2}$ reduction in amplitude; consequently} the residual PSD in Fig.\ref{fig:kagra-psd-k123} {appears} significantly cleaner, improving the sensitivity floor in the {suspension thermal noise-dominated regions}, where some gravitational-wave signals may have measurable SNR.

\renewcommand{\arraystretch}{1.10}
\begin{table}[h!]
\centering
\caption{Kalman bands used to suppress suspension thermal peaks, including central frequency $f_c$ and bandwidth $\Delta f_c$.}
\label{tab:kalman_bands_fc}
\setlength{\tabcolsep}{12pt} % adjust horizontal padding
\begin{tabular}{c r l l} % left-align all columns
\hline
ID & Kalman bands & $f_c$ [Hz] & $\Delta f_c$ [Hz] \\
\hline
1  & 172.0--177.0 Hz & 174.50 & 6 \\
2  & 177.5--188.0 Hz & 182.75 & 8 \\
3  & 402.0--410.0 Hz & 406.00 & 6 \\
4  & 413.0--418.0 Hz & 415.50 & 4 \\
5  & 418.0--424.0 Hz & 421.00 & 6 \\
6  & 424.0--428.5 Hz & 426.25 & 4 \\
7  & 429.0--431.0 Hz & 430.00 & 4 \\
8  & 720.0--722.0 Hz & 721.00 & 4 \\
9  & 726.0--730.0 Hz & 728.00 & 4 \\
10 & 735.0--750.0 Hz & 742.50 & 15 \\
11 & 750.0--770.0 Hz & 760.00 & 20 \\
\hline
\end{tabular}
\end{table}

\begin{figure}[!htb]
    \centering
    \includegraphics[width=\columnwidth]{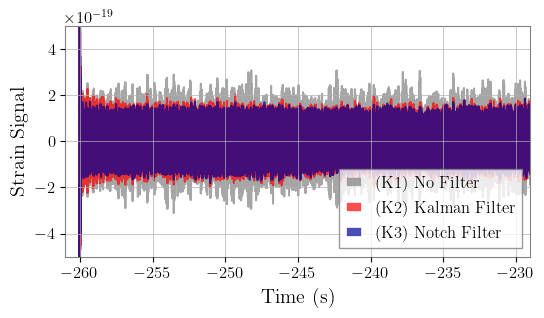}
    \caption{Time series of KAGRA O3GK data starting at GPS 1270312960, shown before and after noise subtraction. The data are bandpassed between 100\,Hz and 1000\,Hz to highlight the suppression of the violin-mode contributions.}
    \label{fig:kagra-ts}
\end{figure}

The signal corrected {using the Kalman filter} in Fig.~\ref{fig:kagra-ts} shows a more stable time distribution as compared to the timeseries of reference within the bandpassed signal between 100Hz--1000Hz. A necessary burn-in time, usually {of a few seconds, is} visible in the first seconds of the time distribution {and} required {for} the Kalman filter to lock {onto} the dynamics of the signal.

%%%%%%%%%%%%%%%%%%%%%%%%%%%%%%%%%%%%%%%%
\subsection{Test Injection into O3GK KAGRA Data}\label{sec:tests}
%%%%%%%%%%%%%%%%%%%%%%%%%%%%%%%%%%%%%%%%

To verify that gravitational-wave signals are preserved, simulated signals must be injected into the strain channel prior to filtering, and the recovered waveforms should be compared after subtraction. From the perspective of gravitational-wave detection, the precise evaluation of any residual contamination is critical and requires further quantification.

%%%%%%%%%%%%%%%%%%%%%%%%%%%%%%%%%%%%%%%%
\subsubsection*{Gravitational Wave Injection}\label{subsec:injection}
%%%%%%%%%%%%%%%%%%%%%%%%%%%%%%%%%%%%%%%%

For the test injection, we {generated} a compact binary coalescence (CBC) waveform using the \textit{SpinTaylorT4} with component masses at 1.4-1.4\,M$_\odot$ and a low-frequency cutoff at 10\,Hz. A distribution of the reference PSD using O3GK was adjusted to {match} the length of the injection template. The
generated waveform {lasts} for about 200s and {is injected} in software into the strain channel. This injection is representative of a binary neutron star merger, producing signals with frequency content overlapping the violin-mode resonances. The template is arbitrarily placed at an SNR of $\rho = 37.67$, as defined by Eq.~(\ref{eq:snr}) using the \textit{sigma} function from PyCBC~\cite{usman2016pycbc}. A broader parameter range and more detailed parameter-estimation studies will be considered in future work to validate the robustness of the subtraction and assess the performance of suspension thermal-noise mitigation across different SNR values.

% \begin{figure}[!htb]
%     \centering
%     \includegraphics[width=\columnwidth]{images/template-bns.png}
%     \caption{Template BNS injection 1.4-1.4\,M$_\odot$ about 200s duration}
%     \label{fig:kagra-psd}
% \end{figure}

%%%%%%%%%%%%%%%%%%%%%%%%%%%%%%%%%%%%%%%%
\subsubsection*{Signal-to-Noise Ratio and $\chi^2$-Significance Tests}\label{subsec:snr}
%%%%%%%%%%%%%%%%%%%%%%%%%%%%%%%%%%%%%%%%

\begin{figure}[!htb]
    \centering
    \includegraphics[width=\columnwidth]{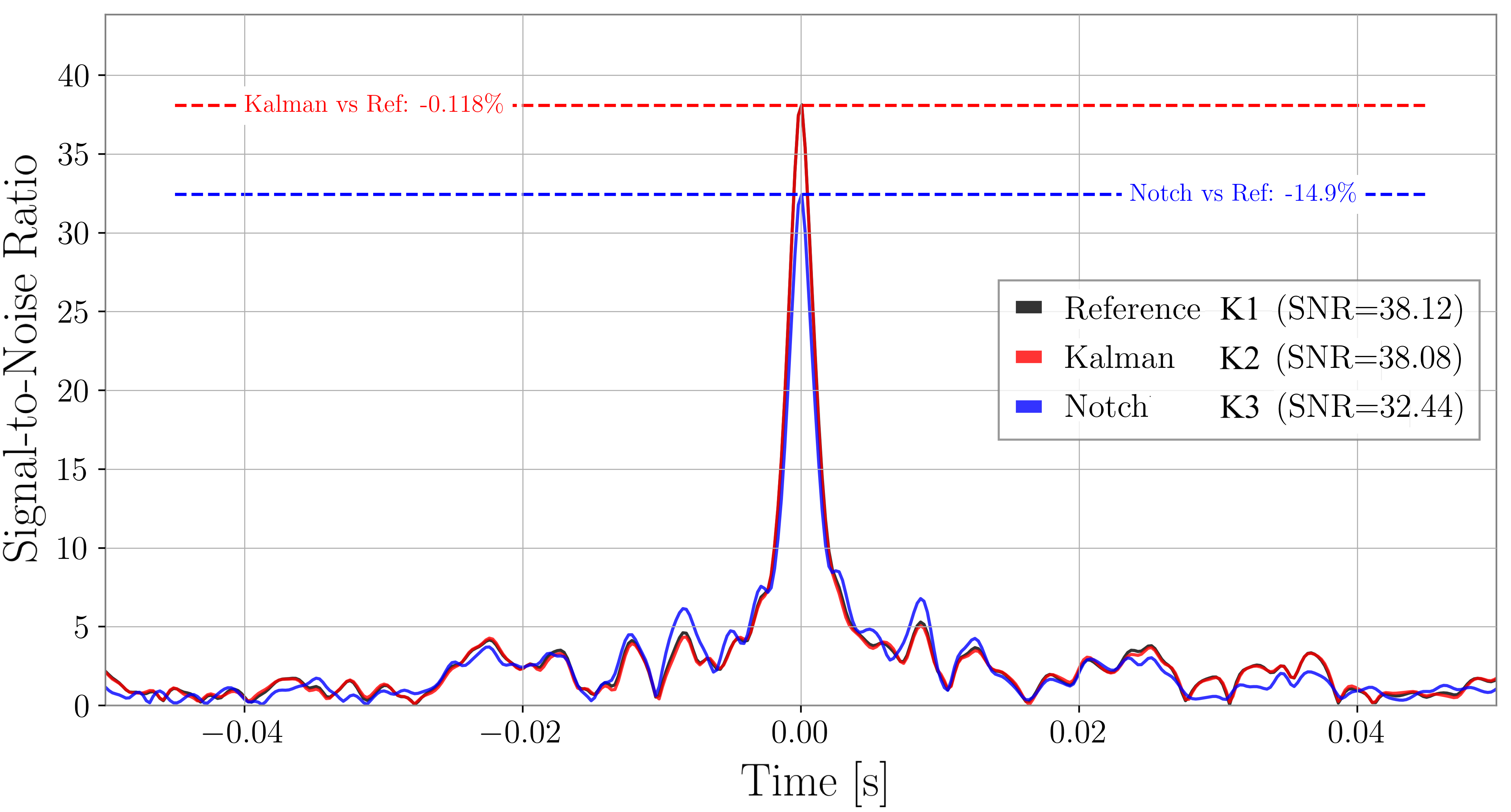}
    \caption{Signal-to-noise ratio distributions obtained from matched filtering using PyCBC~\cite{usman2016pycbc}. $\mathrm{SNR}_{\rm K1}$, $\mathrm{SNR}_{\rm K2}$, $\mathrm{SNR}_{\rm K3}$ are computed against the same PSD reference.}
    \label{fig:kagra-snr}
\end{figure}

For the computation of the signal-to-noise ratio (SNR), denoted $\rho$, as defined in Eq.(\ref{eq:snr}), a reference power spectral density (PSD) is estimated from a 4096\,s segment of O3GK data.
\begin{equation}
\begin{cases}
(s|h)(t) = 4 \, \mathrm{Re} \displaystyle \int 
\frac{\tilde{s}(f) \, \tilde{h}^*(f)}{S_n(f)} \, \mathrm{d}f \\[2mm]
\rho(t) = \dfrac{(s|h)(t)}{(s|s)(t)}
\end{cases}
\label{eq:snr}
\end{equation}
where $S_n(f)$ is the PSD of the reference noise (K1), $s(t)$ is the strain signal, and $h(t)$ is the template waveform. Matched-filter analyses of the injected signals, shown in {Fig.\ref{fig:kagra-snr}, indicate a significant loss in SNR in the notched case (K3) compared to the unfiltered case (K1). 
This occurs because the fraction of the signal lying within the removed bandwidths is fully suppressed by the notch filter, resulting in a drop of the SNR in K3 relative to the reference PSD floor.
}

Considering the frequency dependence and the bandwidth spanned by {the three violin-mode} resonances, a drop of $15\%$ is estimated {as reasonable, knowing that} $\mathrm{SNR}^2 \propto \int_{f_\mathrm{low}}^{f_\mathrm{high}} f^{-7/3}/S_n(f) , df$~\cite{ref:Allen2012,ref:Creighton_2011,ref:Maggiore_2008} and the denoised frequency band{widths}. In the Kalman-filtered case (K2), {no significant drop has been observed} compared to the (K1) reference, suggesting a reasonable subtraction of the violin components within the total signal.

%%%%%%%%%%%%%%%%%%%%%%%%%%%%%%%%%%%%%%%%
%\subsubsection*{$\chi^2$ Significance Test}\label{subsec:chi2}
%%%%%%%%%%%%%%%%%%%%%%%%%%%%%%%%%%%%%%%%

\begin{figure}[!htb]
    \centering
    \includegraphics[width=\columnwidth]{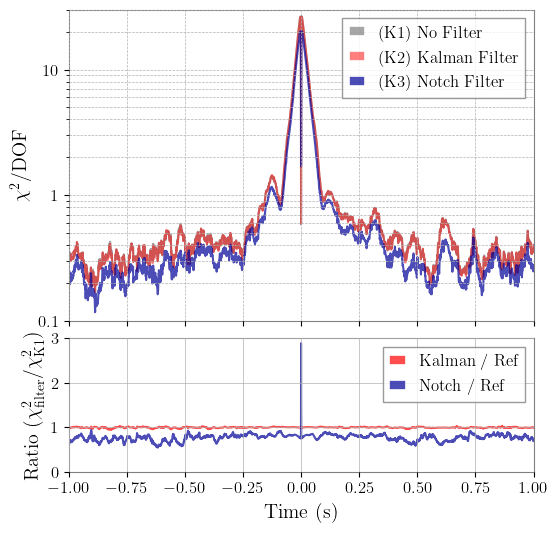}
    \caption{(a) Top: $\chi^2$-test distribution centered on the merger time; {(K1) and (K2) almost perfectly overlap}. (b) Bottom: Ratio plot showing the deviations of (K2) and (K3) relative to the reference signal (K1).}
    \label{fig:kagra-chi2}
\end{figure}

We {performed} the reduced $\chi^2$ test ~\cite{Allen2005,ref:Allen2012,ref:Creighton_2011,ref:Maggiore_2008} on the matched-filter outputs to assess the consistency of the recovered signals with the expected template. The result is shown in Fig.~\ref{fig:kagra-chi2}. The Kalman subtraction does not degrade the $\chi^2$ performance, indicating that the Kalman filter does not introduce systematic distortions into the astrophysical signal band. In Fig.\ref{fig:kagra-chi2}b{, the grey} (K1) and red (K2) curves overlap almost perfectly near $t = 0\,\mathrm{s}$, highlighting the consistency of the residual signal after Kalman filtering. By contrast{,} in Fig.\ref{fig:kagra-chi2}a, a clear $5\sigma$ deviation of the SNR is observed for (K3): the $\chi^2$-distribution decrease sharply to a reasonable value typical from the expected distributions around at $t = 0\,\mathrm{s}$, while (K1) and (K2) remain consistent.

%%%%%%%%%%%%%%%%%%%%%%%%%%%%%%%%%%%%%%%%
\section{Conclusions}\label{sec:concl}
%%%%%%%%%%%%%%%%%%%%%%%%%%%%%%%%%%%%%%%%

We have implemented and validated a Kalman filter approach to remove suspension violin modes from gravitational-wave detector data. Validation against LIGO 40m results confirms that the method reproduces established suppression performance. Applying it to KAGRA O3GK cryogenic data demonstrates its effectiveness in reducing narrow spectral lines while preserving injected gravitational-wave signals.

This first attempt at subtracting suspension thermal noise in a cryogenic interferometer represents an important step toward advancing noise-mitigation strategies for future KAGRA runs. Given the frequency location of the suspension resonances and the $Q$-values estimated in this work, improved control and hardware characterization at KAGRA may provide valuable guidance for suspension-noise mitigation in next-generation observatories such as the Einstein Telescope.

In future work, we aim to strengthen the robustness of the method by finalizing the Kalman framework for real-time implementation in KAGRA’s data stream, with the goal of deployment during O5. Further testing with O4c data will also be conducted to validate the approach under realistic operating conditions.

%%%%%%%%%%%%%%%%%%%%%%%%%%%%%%%%%%%%%%%%
\begin{acknowledgments}
%%%%%%%%%%%%%%%%%%%%%%%%%%%%%%%%%%%%%%%%
This research was supported in part by the Japan Society for the Promotion of Science (JSPS) Grant-in-Aid for Fellows [No. P21726 (M.\ Meyer-Conde)], the JSPS Postdoctoral Fellowships for Research in Japan [No. P25701 (C.\ Alléné)] and the JSPS Grant-in-Aid for Scientific Research  [No. 23K03437 (D.\ Chen)], [No. 22KF0329] (N.\ Kanda)] and [Nos. 23H01176, 23K25872 and 23K22499] (H.\ Takahashi)]. This research was supported by the Joint Research Program of the Institute for Cosmic Ray Research, University of Tokyo, and Tokyo City University Prioritized Studies. We would also thank the LIGO collaboration for sharing the LIGO 40\,m dataset, which is remarkable in its longevity, as data from 1994 continues to provide valuable insights.

% GWOSC: https://gwosc.org/acknowledgement/ (?) 
% Cite: GW OS DATA:
%   KAGRA:2023pio,LIGOScientific:2019lzm,LIGOScientific:2025snk
% This research has made use of data or software obtained from the Gravitational Wave Open Science Center (gwosc.org), a service of the LIGO Scientific Collaboration, the Virgo Collaboration, and KAGRA. This material is based upon work supported by NSF's LIGO Laboratory which is a major facility fully funded by the National Science Foundation, as well as the Science and Technology Facilities Council (STFC) of the United Kingdom, the Max-Planck-Society (MPS), and the State of Niedersachsen/Germany for support of the construction of Advanced LIGO and construction and operation of the GEO600 detector. Additional support for Advanced LIGO was provided by the Australian Research Council. Virgo is funded, through the European Gravitational Observatory (EGO), by the French Centre National de Recherche Scientifique (CNRS), the Italian Istituto Nazionale di Fisica Nucleare (INFN) and the Dutch Nikhef, with contributions by institutions from Belgium, Germany, Greece, Hungary, Ireland, Japan, Monaco, Poland, Portugal, Spain. KAGRA is supported by Ministry of Education, Culture, Sports, Science and Technology (MEXT), Japan Society for the Promotion of Science (JSPS) in Japan; National Research Foundation (NRF) and Ministry of Science and ICT (MSIT) in Korea; Academia Sinica (AS) and National Science and Technology Council (NSTC) in Taiwan.
\end{acknowledgments}

%%%%%%%%%%%%%%%%%%%%%%%%%%%%%%%%%%%%%%%%
\appendix
\section{Resonant Frequency Extractions}\label{app:frequencies}

The values used in Kalman filtering are shown in Table.~\ref{tab:violin-sidebyside}.
The resonant frequencies for the three violin modes are taken from the PTEP \cite{10.1093/ptep/ptac093}, other parameters ($Q_i$ values and $G_i$ values) used in the Lorentzian profile result from a fit. The ${\rm FWHM}_i$ are also shown.

\renewcommand{\arraystretch}{1.05}
\begin{table*}[!htb]
    \centering
    \caption{Fitted parameters of violin modes: $f_0$, $Q$ ($\times10^4$), FWHM ($\times10^3$), and $G_i$ for modes 1--3, shown side by side.}
    \label{tab:violin-sidebyside}
    \begin{tabular}{c c c c c | c c c c c | c c c c c}
        \hline
        \multicolumn{5}{c}{\textbf{Mode 1 (170--186 Hz)}} &
        \multicolumn{5}{c}{\textbf{Mode 2 (402--432 Hz)}} &
        \multicolumn{5}{c}{\textbf{Mode 3 (718--770 Hz)}} \\
        \hline
        & $f_0$ & $Q$ & FWHM & $G_i$ &
        & $f_0$ & $Q$ & FWHM & $G_i$ &
        & $f_0$ & $Q$ & FWHM & $G_i$ \\
        & (Hz) & ($\times 10^4$) & ($\times 10^3$) & &
        & (Hz) & ($\times 10^4$) & ($\times 10^3$) & &
        & (Hz) & ($\times 10^4$) & ($\times 10^3$) & \\
        \hline
 1 & 173.021 & 1.50 & 11.54 & $4.68\times10^{-38}$ &
 1 & 404.961 & 1.61 & 25.09 & $1.65\times10^{-39}$ &
 1 & 721.271 & 1.44 & 49.98 & $1.29\times10^{-40}$ \\
 2 & 174.035 & 1.44 & 12.07 & $9.99\times10^{-39}$ &
 2 & 408.562 & 2.08 & 19.63 & $2.67\times10^{-40}$ &
 2 & 728.291 & 2.79 & 26.10 & $3.42\times10^{-41}$ \\
 3 & 176.510 & 1.46 & 12.12 & $6.41\times10^{-38}$ &
 3 & 414.901 & 1.96 & 21.20 & $1.63\times10^{-39}$ &
 3 & 739.319 & 2.81 & 26.31 & $1.53\times10^{-40}$ \\
 4 & 177.929 & 1.50 & 11.88 & $4.10\times10^{-38}$ &
 4 & 415.482 & 1.87 & 22.23 & $1.94\times10^{-39}$ &
 4 & 743.367 & 1.86 & 40.07 & $7.52\times10^{-41}$ \\
 5 & 178.295 & 1.49 & 11.95 & $1.98\times10^{-38}$ &
 5 & 416.433 & 2.17 & 19.19 & $1.63\times10^{-39}$ &
 5 & 743.762 & 2.79 & 26.65 & $1.25\times10^{-40}$ \\
 6 & 178.387 & 1.50 & 11.89 & $5.25\times10^{-38}$ &
 6 & 416.998 & 1.95 & 21.42 & $7.85\times10^{-40}$ &
 6 & 746.132 & 2.81 & 26.59 & $1.81\times10^{-40}$ \\
 7 & 178.425 & 1.26 & 14.13 & $5.19\times10^{-41}$ &
 7 & 418.844 & 2.15 & 19.48 & $1.43\times10^{-41}$ &
 7 & 746.455 & 2.81 & 26.56 & $1.88\times10^{-40}$ \\
 8 & 178.539 & 1.50 & 11.90 & $6.60\times10^{-38}$ &
 8 & 419.270 & 1.44 & 29.21 & $1.66\times10^{-41}$ &
 8 & 746.707 & 2.80 & 26.68 & $6.12\times10^{-42}$ \\
 9 & 179.052 & 1.42 & 12.65 & $1.05\times10^{-38}$ &
 9 & 419.512 & 1.49 & 28.07 & $3.82\times10^{-40}$ &
 9 & 747.163 & 2.80 & 26.69 & $8.07\times10^{-42}$ \\
10 & 179.247 & 1.10 & 16.23 & $1.33\times10^{-39}$ &
10 & 419.721 & 1.74 & 24.18 & $1.31\times10^{-40}$ &
10 & 749.812 & 2.81 & 26.70 & $4.03\times10^{-41}$ \\
11 & 179.762 & 1.49 & 12.04 & $7.42\times10^{-38}$ &
11 & 419.906 & 1.43 & 29.34 & $3.09\times10^{-40}$ &
11 & 750.900 & 2.80 & 26.82 & $1.47\times10^{-40}$ \\
12 & 179.833 & 1.50 & 11.99 & $4.96\times10^{-38}$ &
12 & 420.296 & 2.17 & 19.37 & $2.65\times10^{-39}$ &
12 & 751.171 & 2.80 & 26.83 & $1.07\times10^{-40}$ \\
13 & 180.061 & 1.50 & 12.02 & $1.26\times10^{-38}$ &
13 & 420.644 & 2.12 & 19.85 & $1.86\times10^{-39}$ &
13 & 751.249 & 2.67 & 28.17 & $2.03\times10^{-40}$ \\
14 & 180.093 & 1.16 & 15.53 & $1.99\times10^{-39}$ &
14 & 421.157 & 2.16 & 19.53 & $8.60\times10^{-40}$ &
14 & 753.344 & 2.80 & 26.94 & $3.78\times10^{-42}$ \\
15 & 180.253 & 1.46 & 12.32 & $6.03\times10^{-38}$ &
15 & 421.269 & 1.82 & 23.13 & $1.24\times10^{-39}$ &
15 & 753.651 & 2.79 & 27.06 & $1.28\times10^{-40}$ \\
16 & 180.625 & 1.49 & 12.11 & $6.56\times10^{-38}$ &
16 & 421.325 & 2.08 & 20.29 & $2.41\times10^{-39}$ &
16 & 754.724 & 2.79 & 27.08 & $7.68\times10^{-41}$ \\
17 & 180.902 & 1.25 & 14.43 & $3.92\times10^{-38}$ &
17 & 421.659 & 2.04 & 20.65 & $1.20\times10^{-39}$ &
17 & 755.729 & 2.80 & 26.99 & $6.82\times10^{-41}$ \\
18 & 180.962 & 1.35 & 13.44 & $2.36\times10^{-38}$ &
18 & 421.744 & 2.01 & 21.03 & $3.95\times10^{-40}$ &
18 & 756.340 & 2.80 & 27.02 & $1.94\times10^{-41}$ \\
19 & 181.615 & 1.46 & 12.46 & $4.68\times10^{-38}$ &
19 & 422.415 & 1.54 & 27.41 & $3.06\times10^{-40}$ &
19 & 756.560 & 2.77 & 27.29 & $1.86\times10^{-41}$ \\
20 & 181.661 & 1.47 & 12.34 & $4.81\times10^{-38}$ &
20 & 422.540 & 2.04 & 20.71 & $2.12\times10^{-39}$ &
20 & 756.750 & 2.80 & 27.03 & $9.82\times10^{-42}$ \\
21 & 182.093 & 1.50 & 12.16 & $9.19\times10^{-40}$ &
21 & 422.642 & 1.81 & 23.31 & $1.13\times10^{-39}$ &
21 & 756.786 & 2.80 & 27.04 & $1.71\times10^{-41}$ \\
22 & 182.139 & 1.44 & 12.67 & $1.42\times10^{-38}$ &
22 & 422.949 & 1.57 & 27.01 & $8.47\times10^{-40}$ &
22 & 756.885 & 2.56 & 29.62 & $2.57\times10^{-40}$ \\
23 & 182.543 & 1.50 & 12.17 & $4.89\times10^{-38}$ &
23 & 424.255 & 1.43 & 29.67 & $3.25\times10^{-41}$ &
23 & 758.686 & 2.71 & 27.95 & $1.31\times10^{-40}$ \\
24 & 182.959 & 1.17 & 15.58 & $2.70\times10^{-39}$ &
24 & 424.658 & 2.15 & 19.79 & $1.12\times10^{-39}$ &
24 & 759.405 & 2.81 & 27.04 & $1.66\times10^{-40}$ \\
25 & 183.427 & 1.50 & 12.27 & $3.47\times10^{-38}$ &
25 & 425.184 & 2.03 & 20.92 & $2.19\times10^{-39}$ &
25 & 760.548 & 2.81 & 27.10 & $6.56\times10^{-41}$ \\
26 & 184.314 & 1.50 & 12.29 & $8.26\times10^{-39}$ &
26 & 425.798 & 2.02 & 21.08 & $1.26\times10^{-39}$ &
26 & 762.359 & 2.80 & 27.23 & $1.01\times10^{-40}$ \\
27 & 184.387 & 1.10 & 16.76 & $8.22\times10^{-39}$ &
27 & 427.256 & 2.07 & 20.67 & $2.14\times10^{-40}$ &
27 & 762.611 & 2.63 & 28.99 & $6.49\times10^{-42}$ \\
28 & N/A & - & - & - &
28 & 427.454 & 2.06 & 20.71 & $1.38\times10^{-39}$ &
28 & 762.930 & 2.77 & 27.55 & $1.08\times10^{-40}$ \\
29 & N/A & - & - & - &
29 & 427.662 & 1.69 & 25.33 & $6.88\times10^{-40}$ &
29 & 764.139 & 2.80 & 27.32 & $5.97\times10^{-42}$ \\
30 & N/A & - & - & - &
30 & 429.796 & 1.86 & 23.14 & $1.92\times10^{-40}$ &
30 & 766.052 & 2.80 & 27.36 & $3.17\times10^{-41}$ \\
31 & N/A & - & - & - &
31 & N/A & - & - & - &
31 & 767.439 & 2.80 & 27.41 & $3.34\times10^{-41}$ \\

32 & N/A & - & - & - &
32 & N/A & - & - & - &
32 & N/A & - & - & - \\

\hline
\end{tabular}
\end{table*}

\section{Derivation of the second-order resonance equations}

Consider a mirror of mass $M$ suspended by four fibers at positions $(x_i, y_i, z_i)$ relative to the mirror’s center of mass (CoM). The mirror has moments of inertia $I_y$ and $I_z$ about the $y$ (pitch) and $z$ (yaw) axes, respectively. Roll effects are neglected, as they have a negligible influence on the measurable interferometer output signal. Each fiber has a mass $m_i$ and stiffness $k_i$. The beam propagates along the $x$ direction, with $y$ being horizontal and $z$ being vertical. The relative elongation of the $i$-th fiber is defined as
\begin{equation}
X_i(t) = x_i(t) - X_{0,i}(t)
\end{equation}
where $x_i(t)$ is the longitudinal displacement of the free end of the fiber and $X_{0,i}(t)$ is the displacement of the corresponding mirror attachment point. For small pitch and yaw angles $\theta_y$ and $\theta_z$, the attachment point motion is
\begin{equation}
X_{0,i}(t) = X(t) - y_i\,\theta_z(t) + z_i\,\theta_y(t)
\end{equation}
with $X(t)$ the CoM translation, and $y_i$, $z_i$ the transverse offsets of the fiber relative to the CoM.  

\medskip

The mirror’s motion couples to the fiber elongations through both translation 
($M\ddot{X}=\sum_i F_{i,x}$) and rotation 
($I_j\ddot{\theta}_j = \tau_j = \sum_i (\mathbf{r}_i \times \mathbf{F}_i)|_j$), leading to
\begin{equation}
\left\{
\begin{aligned}
M \ddot{X} &+ \sum_{i=1}^{4} F_{i,x}(t) = 0 \\
I_z \ddot{\theta}_z &- \sum_{i=1}^{4} y_i\,F_{i,x}(t) = 0 \\
I_y \ddot{\theta}_y &+ \sum_{i=1}^{4} z_i\,F_{i,x}(t) = 0
\end{aligned}
\right.
\end{equation}
For each fiber, the harmonic oscillator equation of the $i$-th fiber follows from Newton’s second law applied to the absolute position $x_i(t)$ of the mass in an inertial frame, with restoring force proportional to the relative elongation $X_i(t) = x_i(t) - X_{0,i}(t)${:}
\begin{equation}
\left\{
\begin{aligned}
m_i\,\ddot{X}_i(t) + k_i\,X_i(t) = -\,m_i\,\ddot{X}_{0,i}(t) \\[6pt]
f_{0,i} = \frac{1}{2\pi}\sqrt{\frac{k_i}{m_i}} \qquad i=1,\ldots,4 
\end{aligned}
\right.
\end{equation}

\noindent
In Fourier space, the relative elongation becomes
\begin{equation}
\tilde{X}_i(\omega)
= \frac{m_i \omega^2}{k_i-m_i\omega^2}\,
\Big[\tilde{X}(\omega)-y_i\tilde{\theta}_z(\omega)+z_i\tilde{\theta}_y(\omega)\Big]
\end{equation}

{The restoring force exerted by fiber $i$ on the mirror is not purely static, but frequency-dependent:}
\begin{equation}
F_{i,x}(\omega) = -\,k_i \tilde{X}_i(\omega) = -\,\frac{k_i m_i \omega^2}{k_i - m_i \omega^2}\tilde{X}_{0,i}(\omega)
\end{equation}
{At frequencies approaching the fiber’s resonance, the inertia of the fiber itself modifies the response of the system. To account for this frequency-dependent effect, we introduce an effective stiffness $\tilde{K}_i(\omega)$:}
\begin{equation}
\tilde{K}_i(\omega) = \frac{k_i m_i \omega^2}{k_i - m_i\omega^2}
\end{equation}

\medskip

The mirror equations of motion in the Fourier domain take the form:
\[
\left\{
\begin{aligned}
-\,M\omega^2 \tilde{X}
  &+ \sum \tilde{K}_i(\omega)\,~~~\bigl[\tilde{X} - y_i \tilde{\theta}_z + z_i \tilde{\theta}_y\bigr] &= 0 \\[4pt]
-\,I_z \omega^2 \tilde{\theta}_z
  &- \sum y_i\,\tilde{K}_i(\omega)\,\bigl[\tilde{X} - y_i \tilde{\theta}_z + z_i \tilde{\theta}_y\bigr] &= 0 \\[4pt]
-\,I_y \omega^2 \tilde{\theta}_y
  &+ \sum z_i\,\tilde{K}_i(\omega)\,\bigl[\tilde{X} - y_i \tilde{\theta}_z + z_i \tilde{\theta}_y\bigr] &= 0
\end{aligned}
\right.
\]

\noindent
These coupled equations can be written compactly in matrix form. In a symmetric suspension geometry around the CoM, the cross-terms vanish so that translational and rotational degrees of freedom decouple. The system then reduces to
\begin{equation}
\underbrace{\begin{bmatrix}
M & 0 & 0 \\
0 & I_z & 0 \\
0 & 0 & I_y
\end{bmatrix}}_{\mathbf{M}}
\begin{bmatrix}
-\,\omega^2 \tilde{X} \\ -\,\omega^2 \tilde{\theta}_z \\ -\,\omega^2 \tilde{\theta}_y
\end{bmatrix}
\;+\;
\underbrace{\begin{bmatrix}
  K_{00} & 0      & 0\\
    0 & K_{yy} & 0 \\
    0 & 0      & K_{zz}
\end{bmatrix}}_{\mathbf{K}(\omega)}
\begin{bmatrix}
\tilde{X} \\[4pt] \tilde{\theta}_z \\[4pt] \tilde{\theta}_y
\end{bmatrix}
= \mathbf{0}
\end{equation}
where
\begin{align*}
K_{00}(\omega) &= \sum_i \tilde{K}_i(\omega), \\
K_{yy}(\omega) &= \sum_i \tilde{K}_i(\omega) y_i^2, \qquad
K_{zz}(\omega) = \sum_i \tilde{K}_i(\omega) z_i^2
\end{align*}

The eigenfrequencies of the suspension modes are then determined by the secular equation
\[
\det\!\left(\mathbf{K}(\omega) - \omega^2 \mathbf{M}\right) = 0
\]

\medskip
\noindent
One eigenmode corresponds to all four fibers oscillating in phase, producing a global center-of-mass displacement of the mirror along the beam axis ($x$) without torque. This translational mode is typically suppressed by the interferometer control system and does not appear in the DARM channel.\\

By contrast, antisymmetric oscillations—such as pairs of fibers stretching in opposite directions—couple directly to the mirror's rotational degrees of freedom. For simplicity purposes, we are also assuming identical fibers ($k_i=k$, $m_i=m$). In this case, the angular eigenvalues describe the yaw and pitch sidebands as
\[
\Delta f_{\rm yaw}   \;\approx\; \frac{f_0}{2 I_z} 
  \sum_{i=1}^{N} m\, y_i^{2},
\qquad 
\Delta f_{\rm pitch} \;\approx\; \frac{f_0}{2 I_y} 
  \sum_{i=1}^{N} m\, z_i^{2}
\]
Here $m$ is the mass of each fiber in its fundamental violin mode; the stiffness $k$ is already absorbed into $f_0$, which is why it does not appear explicitly. Finally, for a cylindrical mirror of radius $R$ and length $L$, the inertia are
\begin{equation}
I_x = \dfrac{1}{2} M R^2, \qquad 
I_y = I_z = \dfrac{M}{12}\left(3R^2+L^2\right)
\end{equation}
Typically, $d_y$ is of order the mirror radius, while $d_z\simeq 0$ when the fibers lie in the horizontal plane. For sapphire fibers and cylindrical test masses, this scaling gives $\Delta f_{\rm yaw}\approx 0.10~\text{Hz}$. Since the actual suspension is not perfectly symmetric (bonding, stiffness, clamp anisotropy, small tension imbalances), this value should be viewed as a lower bound; realistic asymmetries can increase $\Delta f_{\rm yaw}$ into the few-tenths-of-hertz range. Each suspension fiber supports nearly two degenerate violin polarizations: an {x-direction mode} that couples strongly to the interferometer’s arm length readout, and a {y-axis companion} that couples weakly through asymmetries and miscentering. Both modes share the same intrinsic Q, but the yaw-like resonance appears at a lower amplitude and slightly shifted frequency, producing the observed doublets in the violin-mode spectrum.

\section{Process \& Measurement Noise Formula}\label{app:so-formula}

The stochastic dynamics of a suspension mode obey
\begin{equation}
\ddot{x}(t)+\frac{\omega_0}{Q}\,\dot{x}(t)+\omega_0^2 x(t)=\frac{1}{m}\,\xi_x(t)
\end{equation}
with thermal Langevin force \(\xi_x(t)\). By the fluctuation–dissipation theorem~\cite{PhysRev.83.34},
\begin{equation}
\left\{
\begin{aligned}
\langle \xi_x(t)\,\xi_x(t')\rangle &= S_F^{(2)}(\omega_0)\,\delta(t-t')\\[0.4ex]
S_F^{(2)}(\omega_0) &= 2 k_B T\,\mathrm{Re}\left[\,Z(\omega_0)\right] \;=\; 2 k_B T\,m\,\frac{\omega_0}{Q}
\end{aligned}
\right.
\end{equation}
where \(Z(\omega)\) is the mechanical impedance of the oscillator. Because of the delta correlation,  
\[
\langle \xi_x(t)\,\xi_x(t')\rangle=0, \quad \text{for } t\neq t'
\]
\textit{i.e.}, the force is white in time.

After moving to the rotating frame at frequency \(\omega_c\), the effective quadrature drive becomes
\begin{align}
\langle \xi_\psi(t)\,\xi_\psi(t')^\top\rangle
&= \frac{S_F^{(2)}(\omega_0)}{4 m^2 \omega_c^2}\,\delta(t-t')\,\mathbf{I}_2 \notag \\[0.8ex]
&= \frac{k_B T}{2}\,\frac{\omega_0}{Q\,m\,\omega_c^2}\,\delta(t-t')\,\mathbf{I}_2
\end{align}

Discretizing over an effective bandwidth \(\Delta f_c\) around \(\omega_c\), the delta correlation yields a per-sample variance
\begin{equation}
\sigma_{w,i}^2 = \frac{k_B T}{4 \Delta f_c}\,\frac{\omega_0}{Q\,m\,\omega_c^2}
\end{equation}

For the measurement noise, which is dominated by sensing noise rather than thermal driving, a practical estimate can be obtained from the average measured PSD \(\bar P_m\) in the analysis band. Assuming flat noise across \(\Delta f_c\), the corresponding variance is
\begin{equation}
\sigma_v^2 = \frac{1}{\Delta f_c}\,\frac{\bar P_m}{2}
\end{equation}

% \section{Empirical Inharmonic Scaling}\label{app:inharmonic}

% In the simplest case of harmonic oscillators, the frequencies of higher-order harmonics depend only on the fundamental frequency $f_0$ scaled by the mode number $n$ : $f_n = nf_0$.
% This proportional scaling can be observed in the LIGO 40m prototype until the 8th violin modes order (see Figure \ref{app:fig:ligo40m}).

\begin{figure*}[!htb]
    \centering
    \includegraphics[width=2\columnwidth]{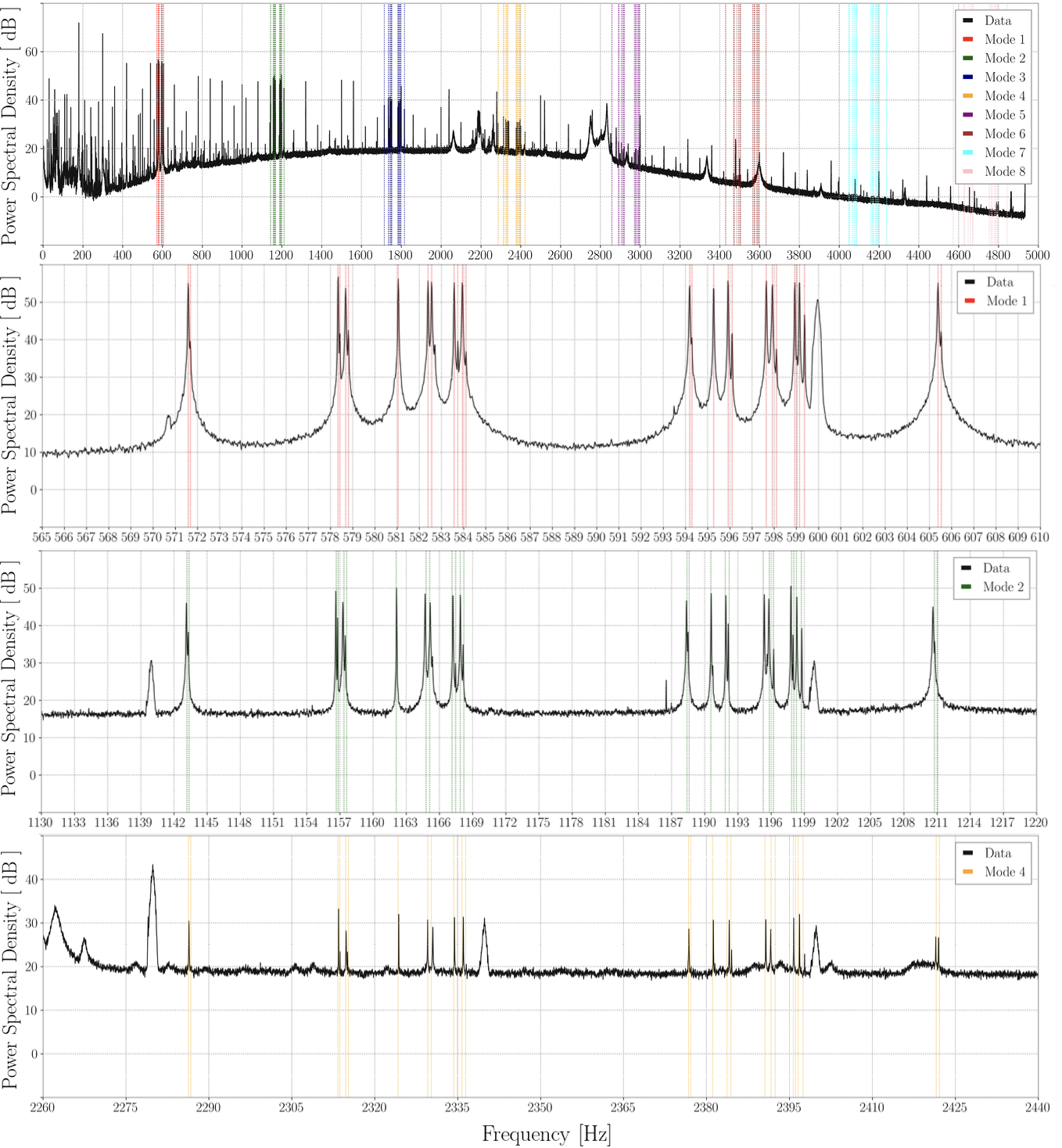}
    \caption{(a) Top: LIGO 40m PSD. The colored vertical lines show the violin modes frequencies for each of the 8th first modes. (b), (c) and (d) Respectively, the 1st, 2nd and 4th violin modes in LIGO 40m.}
    \label{fig:ligo40m-psd-modes}
\end{figure*}

\bibliography{main}
\end{document}